\documentclass[twocolumn,floats,floatfix,superscriptaddress]{revtex4-2}

\usepackage{amsbsy}
\usepackage{amssymb}
\usepackage{amsmath}
\input{epsf}
\usepackage{graphicx}
\usepackage{color}

\usepackage[caption=false]{subfig}
\usepackage{float}
\usepackage{subfig}
\usepackage{gensymb}
\usepackage{units}
\usepackage[flushleft]{threeparttable}
\usepackage[utf8]{inputenc}
\usepackage{aas_macros}
\usepackage{subfig}
\usepackage{lipsum}

\usepackage[final]{changes}

\definechangesauthor[name=Alan, color=blue]{0}
\definechangesauthor[name=Xiaoyi, color=red]{1}

\usepackage{tikz,xcolor,hyperref}

\definecolor{lime}{HTML}{A6CE39}
\DeclareRobustCommand{\orcidicon}{%
	\begin{tikzpicture}
	\draw[lime, fill=lime] (0,0) 
	circle [radius=0.16] 
	node[white] {{\fontfamily{qag}\selectfont \tiny ID}};
	\draw[white, fill=white] (-0.0625,0.095) 
	circle [radius=0.007];
	\end{tikzpicture}
	\hspace{-2mm}
}

\foreach \x in {A, ..., Z}{%
	\expandafter\xdef\csname orcid\x\endcsname{\noexpand\href{https://orcid.org/\csname orcidauthor\x\endcsname}{\noexpand\orcidicon}}
}

\begin{document}
\title{Bridging relativistic jets from black hole scales to long-term electromagnetic radiation distances: A moving-mesh general relativistic hydrodynamics code with the HLLC Riemann solver}

\author{Xiaoyi Xie \orcidA{}}
\email{xxie@aei.mpg.de}
\affiliation{%
  Max Planck Institute for Gravitational Physics (Albert Einstein Institute), D-14476 Potsdam, Germany
}%
\author{Alan Tsz-Lok Lam \orcidB{}}%
\affiliation{%
  Max Planck Institute for Gravitational Physics (Albert Einstein Institute), D-14476 Potsdam, Germany
}%




\begin{abstract}

Relativistic jets accompany the collapse of massive stars, the merger of compact objects, or the accretion of gas in active galactic nuclei. They carry information about the central engine and generate electromagnetic radiation. No self-consistent simulations have been able to follow these jets from their birth at the black hole scale to the Newtonian dissipation phase, making the inference of central engine property through astronomical observations undetermined. We present the general relativistic moving-mesh framework to achieve the continuity of jet simulations throughout space and time. We implement the general relativistic extension for the moving-mesh relativistic hydrodynamic code, \texttt{JET}, and develop a tetrad formulation to utilize the Harten–Lax–van Leer Contact (HLLC) Riemann solver in the general relativistic moving-mesh code. The new framework is able to trace the radial movement of relativistic jets from central regions where strong gravity holds all the way to distances of jet dissipation. 
\end{abstract}


\keywords{General Relativistic, Relativistic jet, HLLC}
\maketitle

\section{Introduction}

Relativistic collimated outflows, known as jets, are associated with many astrophysical systems of vastly different scales, from stellar to galactic and even to extra-galactic levels. Phenomena like microquasars, young stellar objects, gamma-ray bursts (GRBs), active galactic nuclei (AGN), and quasars demonstrate the prevalence of relativistic jets and highlight the ubiquity of the underlying physical processes that give rise to these phenomena.

A central aspect shared by these varied astrophysical systems is the phenomenon of accretion, in which matter is attracted and pulled into a dense celestial body, like a black hole or neutron star. As matter falls onto these objects, gravitational and magnetic forces play crucial roles in launching and collimating the relativistic jets. Studying relativistic jets across different scales provides astronomers with a unique opportunity to probe fundamental astrophysical processes and test our understanding of high-energy physics in extreme environments. 

Commencing with the Penrose process \cite{1969NCimR...1..252P,1971NPhS..229..177P}, numerous theoretical investigations have been undertaken to explore jets and mass outflows near black holes. The Penrose process initially elucidates energy extraction from in-falling matter into a rotating black hole. Subsequently, the seminal work by \citeauthor{1977MNRAS.179..433B} (BZ) demonstrated that jet energy could be extracted from the rotational energy of large-scale magnetic fields surrounding spinning black holes. Later, \citeauthor{1982MNRAS.199..883B} (BP) highlighted that matter could also depart from the surface of the accretion disk due to magneto-centrifugal acceleration. 

One of the fundamental questions in accretion disk physics is how the angular momentum transfers within the disk. Initially, \citeauthor{1973A&A....24..337S} introduced the '$\alpha$-disc' model in a groundbreaking paper. However, the source of the ad hoc viscosity in this model remains questionable. In contrast, recent years have seen widespread acceptance of magneto-rotational instability (MRI; \citeauthor{1991ApJ...376..214B}) as the primary mechanism for angular momentum transport in accretion flows. 

Another fundamental question in accretion disk physics is the generation of  the large poloidal magnetic field as it is pretty natural to assume a toroidal field configuration for accretion flows. To begin with, the orbital differential shear would predominantly amplify the toroidal magnetic field by the shearing of seed poloidal magnetic field, the so-called $\Omega$ effect. It took simulators many years to achieve the necessary resolutions and finally report the self generation of the large-scale poloidal magnetic field in black hole accretion disk due to the $\alpha$-effect (which relies on the buoyancy and Coriolis forces to convert toroidal into poloidal magnetic flux) \cite{2020MNRAS.494.3656L,2021NewAR..9201610K}. The general mean-field dynamo theory (see, e.g., \cite{1978mfge.book.....M,1980mfmd.book.....K,1981ARA&A..19..115C,doi:10.1146/annurev.fl.24.010192.002331,BRANDENBURG20051}) has been widely used to investigate the generation of large-scale magnetic fields from small-scale turbulence.

Recent long-term general-relativistic (GR) neutrino-radiation magnetohydrodynamics (MHD) simulations of the merger of the binary neutron star and black hole neutron star have shown that effective viscous processes, $\alpha-\Omega$ magnetic dynamo can lead to the generation of large-scale magnetic field, and post-merger mass ejection \cite{2023PhRvL.131a1401K,2024NatAs...8..298K,2023PhRvD.107l3001H}. The analysis of the binary neutron star (BNS) merger remnant and post-merger ejecta has been investigated in detail (see, e.g., \cite{2018ApJ...860...64F,2021ApJ...913..100K,2023ApJ...942...39F}). Still, the process of successfully launching a relativistic jet is undoubtedly complex. For a comprehensive understanding of the launching mechanism, general relativistic magnetohydrodynamic (GRMHD) simulations that integrate intricate microphysical processes are imperative. On the other hand, relativistic outflows  play a pivotal role in a multitude of astronomical phenomena. For example, it has been speculated that the BNS merger remnants and relativistic ejecta are the central engines of gamma-ray bursts \cite{1989Natur.340..126E,1993ApJ...405..273W,2004RvMP...76.1143P,2015PhR...561....1K} and kilo-nova \cite{1998ApJ...507L..59L,2010MNRAS.406.2650M,2011ApJ...738L..32G,2015MNRAS.450.1777K,2018ApJ...869..130R,2019LRR....23....1M}. Relativistic outflows or jets are instrumental in shaping the emission profiles and contributing significantly to the high-energy radiation observed. Understanding these electromagnetic observations requires tracking the propagation of relativistic jets and their interaction with the ambient medium for a long period of time. However, simulating the complete journey of relativistic jets and the related emission process is numerically challenging. Studies in literature split focus on various parts of the whole process. Many studies conduct MHD/GRMHD simulations to investigate the jet launching process and early propagation (see, e.g., \cite{1998ApJ...495L..63K,1999ApJ...522..727K,2002Sci...295.1688K,2004ApJ...615..389M,2005A&A...436..273A,2005A&A...436..273A,2006MNRAS.368.1561M,2009MNRAS.396.2038B,2021MNRAS.502.1843N,2011MNRAS.418L..79T,2021ApJ...915L...4G,2022NatAs...6..103C,2022ApJ...933L...9G}). Some other studies use special relativistic MHD/HD simulations to investigate the jet's interaction with the ambient medium, away from the central compact region (see, e.g., \cite{1994ApJ...436L.119D,1994A&A...281L...9M,1997ApJ...479..151M,1998MNRAS.297.1087K,2004ApJ...608..365Z,2017ApJ...835L..34M,2018MNRAS.475.2971B,2018MNRAS.478.4553D,2018ApJ...866....3D,2018ApJ...863...58X,2000ApJ...531L.119A,2021MNRAS.500..627H}). In this study, we propose a formulation to achieve the continuum of jet simulations throughout space and time and potentially bridge these two research domains. The formulation is built upon the development of the moving-mesh technique \cite{2010MNRAS.401..791S,2011ApJS..197...15D,2012ApJ...758..103G,2013ApJ...775...87D,2015ApJS..216...35Y,2016A&C....16..109V,2016A&C....16..109V,2016A&C....16..109V,2020MNRAS.496..206C,2022MNRAS.510.1315A,10.1093/mnras/stae057}, which has demonstrated its efficiency in simulating ultrarelativistic jets (see, e.g., \cite{2015ApJ...806..205D,2018ApJ...863...58X,2019ApJ...880..135X}). The extension of the moving-mesh technique to the general relativistic hydrodynamics only appears in recent years. We have seen several moving-mesh codes been extended to include GR effects \cite{2017ApJ...835..199R,2020MNRAS.496..206C,10.1093/mnras/stae057}. Most of these moving-mesh codes use Harten–Lax–van Leer (HLL) or Harten-Lax-van Leer-Einfeldt (HLLE) Riemann solver \cite{1983JCoPh..49..357H,1988SJNA...25..294E}. However, the HLLC approximate Riemann solver \cite{2005MNRAS.364..126M} resolves not only the extremal waves but also the contact discontinuity in the Riemann fan and is useful for maintaining contact discontinuities with high precision. Its implementation in fixed-mesh GR employs a local frame transformation\cite{2016ApJS..225...22W,2022PhRvD.106l4041K}. In this study, we provide the mathematical formulation of incorporating the HLLC Riemann solver into a general relativistic moving-mesh code and demonstrate its robustness in simulating fluid flows under strong gravity. In Sec. \ref{sec:refmetric}, we implement the general relativistic extension to the special relativistic moving-mesh hydrodynamic code \textit{JET} \cite{2013ApJ...775...87D} using the reference metric formulation \cite{PhysRevD.79.104029,2012PhRvD..85l4037M,2013PhRvD..87d4026B,2014PhRvD..89h4043M}. In Sec. \ref{sec:tetrad}, we illustrate the tetrad formulation for solving the HLLC Riemann problem in general relativity and the procedures to incorporate it into the moving-mesh framework. Section \ref{sec:techniques} presents several code implementation techniques. In Sec. \ref{sec:fixed_mesh_test}, we conduct several simulations with fixed mesh to test the robustness of the GR extension in the code. In Sec. \ref{sec:moving_mesh_test}, we conduct numerical tests with the moving-mesh grid demonstrating the code's capability to track and resolve the relativistic outflow. For the first time in literature, we successfully launch a relativistic jet from the black hole-torus system and simulate its complete propagation to the dissipation distance. Such simulation provides additional evidence supporting the feasibility of full-time-domain jet simulations, as discussed in our earlier research \cite{2019ApJ...880..135X}. Conclusions and future work are discussed in Sec. \ref{sec:conclusion}.

Throughout this paper, we use the Greek indices $\left( \alpha, \beta, \mu, \nu, \dots \right)$ running from $0$ to $3$ to denote the spacetime components, and the Latin indices $\left( i,j,k, \dots \right)$ running from $1$ to $3$ to denote the space components. We adopt the geometric units $G = c = M_{\odot}= 1$ throughout this paper. All the length scales and timescales are expressed in units of the gravitational radius $r_g = G M_{\odot}/c^2$ and $t_g = r_g/c$, respectively, unless stated otherwise.

\section{General relativistic hydrodynamics in a reference metric formulation}\label{sec:refmetric}

The 2D special relativistic moving-mesh hydrodynamic code \textit{JET} adopts spherical coordinates assuming axisymmetry. The cell interfaces orthogonal to the radial direction are allowed to move radially. The code is essentially Lagrangian in the radial direction, coupled laterally by transverse flux. This setup is particularly suitable for modeling relativistic radial outflows \cite{2011ApJS..197...15D}. To minimize the modifications for the code, we derive the general relativistic hydrodynamic equations in a way that resembles the special relativistic counterparts. 
In the following, we lay out the implementation steps for clarity. Despite of the axisymmetry property of the \textit{JET} code, throughout this paper we will show all the derivations without imposing any symmetry for completeness.

In the standard 3+1 decomposition (see, e.g., \cite{10.1093/acprof:oso/9780199205677.001.0001,2011PhT....64b..49B,shibata2015numerical}), the spacetime is foliated by a family of spatial hypersurface $\Sigma_t$ with future-pointing timelike unit normal vector denoted by $n^\mu$,
which decomposes the line element as
\begin{align}
    ds^2 &= -\alpha^2 dt^2 + \gamma_{ij}(dx^i+\beta^i dt)(dx^j + \beta^j dt)
\end{align}
where $\alpha$ is the lapse function, $\beta^i$ is the shift vector, and $\gamma_{ij}$ is the spatial metric induced on $\Sigma_t$.
In terms of the lapse and shift, the normal vector $n^\mu$ can be expressed as 
\begin{align}
    n_\mu &= (-\alpha, 0, 0, 0), & n^\mu &= (1/\alpha, -\beta^i/\alpha).
\end{align}

We adopt a conformal decomposition of the spatial metric $\gamma_{ij}$
\begin{align} \label{eq:conformal_scaling}
    \gamma_{ij} &= \psi^4 \bar{\gamma}_{ij}, & \gamma^{ij} &= \psi^{-4} \bar{\gamma}^{ij}
\end{align}
where $\psi := (\gamma/\bar{\gamma})^{1/12}$ is the conformal factor, $\bar\gamma_{ij}$ is the conformal spatial metric, and $\gamma$ and $\bar\gamma$ are the determinants of $\gamma_{ij}$ and $\bar\gamma_{ij}$ respectively.
Following the reference-metric formulation (as shown in \cite{2018PhRvD..97h4059M}), we define the residual metric $\epsilon_{ij}$ as
\begin{equation}
    \epsilon_{ij} := \bar{\gamma}_{ij} - \hat{\gamma}_{ij},
\end{equation}
where $\hat\gamma_{ij}$ is a time-independent background reference metric.
For our purpose, we specialize $\hat\gamma_{ij}$ to be a flat metric in spherical coordinate $(r, \theta, \phi)$
as $\hat{\gamma}_{ij} := {\rm diag}(1, r^2, r^2\sin^2 \theta)$.
To make the conformal scaling unique, we set $\bar{\gamma} = \hat{\gamma} := \det(\hat{\gamma}_{ij})$ (see, e.g., \cite{2018PhRvD..97f4036R}). We denote $\nabla_\mu$, $\mathcal{D}_i$, $\bar{ \mathcal{D}}_i$ and $\hat{\mathcal{D}}_i$ as the covariant derivatives of
spacetime metric $g_{\mu\nu}$, $\gamma_{ij}$, $\bar\gamma_{ij}$ and $\hat\gamma_{ij}$ respectively.

The equations of relativistic hydrodynamics are based on conservation of rest mass
\begin{equation}\label{eq:conservation_mass}
    \nabla_\mu(\rho u^\mu) = 0,
\end{equation}
and conservation of energy-momentum
\begin{equation}\label{eq:conservation_energy_momentum}
    \nabla_\nu(T^{\mu\nu}) = 0,
\end{equation}
where $\rho$ is the rest-mass density and $u^\mu$ is the fluid four-velocity and $T^{\mu\nu}$ is the stress-energy tensor.
Here we assume perfect fluid for $T^{\mu\nu}$ in the form
\begin{equation}
    T^{\mu\nu}= \rho h u^\mu u^\nu + P g^{\mu \nu},
\end{equation}
where $P$ is the pressure, $\varepsilon$ is the specific internal energy and $h:=1+\varepsilon + P/\rho$ is the specific enthalpy. 
In 3+1 decomposition, $T^{\mu\nu}$ can be decomposed as
\begin{subequations}
\begin{align}
    S_0 &:= n_\mu n_\nu T^{\mu\nu} &&= \rho h W^2 -P, \\
    S^j &:= - \gamma_\mu^i n_\nu T^{\mu\nu} &&= \rho h  W^2 v^i, \\
    S^{ij} &:= \gamma_\mu^i \gamma_\nu^j T^{\mu\nu} &&= \rho h W^2 v^i v^j + P \gamma^{ij},
\end{align}
\end{subequations}
where $W:= \alpha u^t$ is the Lorentz factor and $v^i := u^i /W + \beta^i / \alpha$ is the fluid velocity measured by the normal observer.

We adopt the Valencia formulation in reference metric formulation following \cite{2012PhRvD..85l4037M,2020PhRvD.101j4007M}
to rewrite the hydrodynamics equations in conservative form as
\begin{equation}\label{eq:form_eq}
    \partial_{t} \boldsymbol{q}+\frac{1}{\sqrt{\hat{\gamma}}} \partial_{j}\left[\sqrt{\hat{\gamma}} \boldsymbol{f}^{j}\right]=\boldsymbol{s}
\end{equation}
with state vectors $\boldsymbol{q}$ being the conserved variables
\begin{align} \label{eq:q_def}
    \boldsymbol{q} &=
    \begin{pmatrix}
    q_{D} \\
    q_{S_i} \\
    q_{\tau}
    \end{pmatrix} =
    \psi^6 \sqrt{\bar{\gamma}/\hat{\gamma}}
    \begin{pmatrix}
    D \\
    S_i \\
    \tau
    \end{pmatrix}, 
\end{align}
where $(D, S_i, \tau) := (\rho W, \rho h W u_i, \rho h W^2 - P - D)$
are the density, momentum density and energy density variables in Valencia form respectively.
$\boldsymbol{f}^{j}$ and $\boldsymbol{s}$ represent the flux and source terms respectively written as
\begin{align} 
    \boldsymbol{f}^{j} &=
    \begin{pmatrix}
    \left( f_D \right)^j \\
    \left( f_{S_i} \right)^j \\
    \left( f_\tau \right)^j
    \end{pmatrix}, \quad
    \boldsymbol{s} =
    \begin{pmatrix}
    0 \\
    s_{S_i} \\
    s_\tau
    \end{pmatrix}.
\end{align}
The detailed derivation is shown in Appendix~\ref{append:gr_equations} for the readers' interests.

One key ingredient of the reference metric method is to evolve tensorial quantities in an orthonormal basis with respect to the background metric. In this way, all tensor components are explicitly free of coordinate singularities. We will follow the notation of \cite{2020PhRvD.101j4007M} to distinguish between coordinate-basis and orthonormal-basis components. The plain Latin indices represent the tensor components in the standard coordinate basis, while the Latin indices surrounded with curly braces denote the components in the background orthonormal basis. We also introduce a set of basis vector $\hat{e}^{\{k\}}_i$ that are orthonormal with respect to the background metric $\hat{\gamma}_{ij}$,
\begin{equation}
\hat{\gamma}_{i j}=\delta_{\{k\}\{l\}} \hat{\mathbf{e}}_i^{\{k\}} \hat{\mathbf{e}}_j^{\{l\}} .
\end{equation}

For the flat background metric in spherical coordinates, this leads to
\begin{eqnarray}
  \hat{\mathbf{e}}_i^{\{k\}} &=& {\rm{diag}}(1,r,r \sin \theta), \\
  \hat{\mathbf{e}}^k_{\{i\}} &=&  {\rm{diag}}(1, 1/r, 1/(r \sin \theta)).
\end{eqnarray}

 So any tensor $A^i{}_j$ defined in the standard coordinate basis can be decomposed into its orthonormal basis counterpart $A^{\{i\}}{ }_{\{j\}}$ as
\begin{equation}
    A^i{}_j = A^{\{k\}}{}_{\{l\}} \hat{\mathbf{e}}^i_{\{k\}} \hat{\mathbf{e}}_j^{\{l\}}.
\end{equation}

As an example, the residual metric $\epsilon_{ij}$ can be expressed in terms of the components in the orthonormal basis $ \hat{\mathbf{e}}^{\{k\}}_i$ as
\begin{align}
\epsilon_{i j} &=
\begin{pmatrix}
\epsilon_{\{r\}\{r\}} & r \epsilon_{\{r\} \{\theta\}} & r \sin \theta \epsilon_{\{r\} \{\phi\}} \\
r \epsilon_{\{r\} \{\theta\}} & r^2 \epsilon_{\{\theta\} \{\theta\}} & r^2 \sin \theta \epsilon_{\{\theta\} \{\phi\}} \\
r \sin \theta \epsilon_{\{r\} \{\phi\}} & r^2 \sin \theta \epsilon_{\{\theta\} \{\phi\}} & r^2 \sin ^2 \theta \epsilon_{\{\phi\} \{\phi\}}
\end{pmatrix},
\end{align}
while for the conserved momentum we have
$(q_{S_r}, q_{S_\theta}, q_{S_\phi}) = \psi^6 \sqrt{\bar{\gamma}/\hat{\gamma}} \left(S_{\{r\}}, r S_{\{\theta\}}, r\sin\theta S_{\{\phi\}} \right)$.

The complete set of general relativistic hydrodynamic equations in 3D spherical coordinates under reference metric formalism (\ref{eq:form_eq}) can be derived as:

\begin{widetext}
\begin{subequations}\label{eq:complete_eqs}
\begin{align} 
\begin{split}
    \frac{\partial}{\partial t} \left( \mathcal{M} D \right) 
    &+ \frac{1}{r^2} \frac{\partial}{\partial r} \left[r^2 \alpha\mathcal{M} D \bar{v}^{\{r\}}\right] 
    + \frac{1}{r \sin \theta} \frac{\partial}{\partial \theta} \left[ \sin\theta \alpha \mathcal{M} D \bar{v}^{\{\theta\}}\right] 
    + \frac{1}{r\sin \theta} \frac{\partial}{\partial \phi} \left[ \alpha \mathcal{M}  D \bar{v}^{\{\phi\}} \right] =0
\end{split} \\
\begin{split}
    \frac{\partial}{\partial t} \left( \mathcal{M} S_{\{r\}} \right) 
    &+ \frac{1}{r^2} \frac{\partial}{\partial r} \left[r^2\left(\alpha \mathcal{M} (S_{\{r\}} \bar{v}^{\{r\}}+ P )\right) \right] 
    + \frac{1}{r\sin \theta} \frac{\partial}{\partial \theta} \left[ \sin \theta \left( \alpha \mathcal{M} S_{\{r\}} \bar{v}^{\{\theta\}} \right) \right] 
    + \frac{1}{r \sin \theta} \frac{\partial}{\partial \phi} \left[\alpha \mathcal{M} S_{\{ r\}} \bar{v}^{\{\phi \}} \right] \\ 
    &= \mathcal{M} \left\{ \alpha \left[
    \frac{2P}{r} \left(\frac{\bar{\gamma}^{\{\theta\}\{\theta\}} + \bar{\gamma}^{\{\phi\}\{\phi\}}}{2} \right) 
    + \frac{\psi^4 \rho h W^2}{r} \left( v^{\{\theta\} 2} + v^{\{\phi\} 2} \right) 
    + \frac{1}{2} \psi^4 S^{jk}\partial_r \epsilon_{jk}
    \right. \right. \\
    & \left.\left.
    + 2 (\rho h (W^2-1) + 3P) \partial_r \ln\psi  \right]  
    -  S_0 \partial_r \alpha 
    + S_{\{i\}} \partial_r \beta^{\{i\}}
    - \frac{1}{r} \left( S_{\{\theta\}} \beta^{\{\theta\}} + S_{\{\phi\}}\beta^{\{\phi\}} \right) \right\}
\end{split} \\
\begin{split}
    \frac{\partial}{\partial t} \left(r \mathcal{M} S_{\{\theta\}} \right) 
    &+ \frac{1}{r^2} \frac{\partial}{\partial r} \left[ r^2 \alpha \left(  r \mathcal{M} S_{\{\theta\}} \right) \bar{v}^r  \right] 
    + \frac{1}{r \sin \theta} \frac{\partial}{\partial \theta} \left[ r \sin \theta \alpha \mathcal{M} \left( S_{\{\theta\}}\bar{v}^{\{\theta\}}+ P \right)\right] 
    + \frac{1}{r \sin\theta} \frac{\partial}{\partial \phi} \left[ r \alpha \mathcal{M} S_{\{\theta\}} \bar{v}^{\{\phi\}} \right] \\
    &= \mathcal{M} \left\{ \alpha \left[ 
    \left(P \bar{\gamma}^{\{\phi\}\{\phi\}} + \psi^4 \rho h W^2 v^{\{\phi\}2} \right) \cot \theta 
    + \frac{1}{2} \psi^4 S^{jk}\partial_\theta \epsilon_{jk} 
    + 2 (\rho h(W^2-1) + 3P) \partial_\theta \ln \psi 
    \right] \right. \\
    &\left. - S_0 \partial_\theta \alpha 
    + S_{\{i\}} \partial_\theta \beta^{\{i\}} 
    - S_{\{\phi\}} \beta^{\{\phi\} }\rm{cot}\theta \right\}
\end{split} \\
\begin{split}
    \frac{\partial}{\partial t}(r \sin \theta &\mathcal{M} S_{\{\phi\}}) 
    + \frac{1}{r^2} \frac{\partial}{\partial r} \left[ r^2 \alpha (r \sin \theta \mathcal{M} S_{\{\phi\}}) \bar{v}^r \right] 
    + \frac{1}{r\sin \theta} \frac{\partial}{\partial \theta} \left[\sin \theta \alpha (r\sin \theta \mathcal{M} S_{\{\phi\}}) \bar{v}^{\{\theta\}} \right] \\
    &+ \frac{1}{r\sin \theta} \frac{\partial}{\partial \phi} \left[ r \sin \theta \alpha \mathcal{M} (S_{\{\phi\}} \bar{v}^{\{\phi\}} + P)
    \right] \\
    &= \mathcal{M} \left\{\alpha \left[ 2 (\rho h(W^2-1) + 3P) \partial_\phi \ln \psi 
    +  \frac{1}{2} \psi^4 S^{jk} \partial_\phi \epsilon_{jk} \right] 
    - S_0\partial_\phi \alpha 
    + S_{\{i\}} \partial_\phi \beta^{\{i\}} \right\} 
\end{split} \\
\begin{split}
    \frac{\partial}{\partial_t} (\mathcal{M} \tau) 
    &+ \frac{1}{r^2} \frac{\partial}{\partial r} \left[ r^2 \alpha \mathcal{M} \left( \tau \bar{v}^{\{r\}} +  P v^{\{r\}} \right) \right] 
    + \frac{1}{r\sin\theta} \frac{\partial}{\partial \theta} \left[ \sin \theta \alpha\mathcal{M} \left( \tau \bar{v}^{\{\theta\}} + P \,v^{\{\theta\}} \right) \right] \\
    &+ \frac{1}{r\sin\theta} \frac{\partial}{\partial \phi} \left[ \alpha \mathcal{M} \left( \tau \bar{v}^{\{\phi\}} +  P v^{\{\phi\}} \right) \right]  \\
    &= \alpha \mathcal{M} \left[
    T^{00}(\beta^i\beta^j K_{ij} - \beta^i\partial_i \alpha) + T^{0i} (2\beta^jK_{ij} - \partial_i \alpha) + T^{ij}K_{ij} \right] 
\end{split}
\end{align}
\end{subequations}
\end{widetext}
where $K_{ij}$ is the extrinsic curvature,
$\mathcal{M}:=\psi^6 \sqrt{\bar{\gamma}/\hat{\gamma}}$
and $\bar v^{\{i\}} := v^{\{i\}} - \beta^{\{i\}} / \alpha$, 
alongside with the special relativistic Eq.~(\ref{eq:complete_eqs_sr}) in Appendix~\ref{app:hydro_eq_sr} for comparison
(see also \cite{2006ApJS..164..255Z}).
Noted that in practice we compute $\partial_k \epsilon_{\{i\}\{j\}}$ numerically in source terms instead of $\partial_k \epsilon_{ij}$ as
\begin{equation}
  \partial_k \epsilon_{i j}=\hat{\mathbf{e}}_i^{\{l\}} \hat{\mathbf{e}}_j^{\{m\}} \partial_k \epsilon_{\{l\}\{m\}}+\epsilon_{\{l\}\{m\}} \partial_k\left(\hat{\mathbf{e}}_i^{\{l\}} \hat{\mathbf{e}}_j^{\{m\}}\right),
\end{equation}
while the second term $\partial_k\left(\hat{\mathbf{e}}_i^{\{l\}} \hat{\mathbf{e}}_j^{\{m\}}\right)$ is evaluated analytically.

\section{Tetrad Formation and the HLLC Riemann solver}\label{sec:tetrad}

To evaluate the numerical flux through cell interfaces, HLL-type (HLLE/HLLC) Riemann solvers have been designed for relativistic hydrodynamics in Minkowski spacetime \cite{2005MNRAS.364..126M,1994ShWav...4...25T}. Most of the GRHD/GRMHD codes in the literature use HLLE Riemann solver in curved spacetime (see, e.g. \cite{2005PhRvD..72d4014S,2000LRR.....3....2F,2021MNRAS.508.2279C,2003ApJ...589..444G}). The HLLC Riemann solver that captures the contact discontinuity in the wave fan has recently been added for GR codes \cite{2016ApJS..225...22W,2022PhRvD.106l4041K,2011A&A...528A.101B}. We follow previous works for the implementation of the HLLC Riemann solver in general relativity \cite{1998A&A...339..638P,2016ApJS..225...22W,2016ApJS..225...22W,2022PhRvD.106l4041K}. The basic idea is based on the equivalence principle: physical laws in a local inertial frame of a curved spacetime have the same form as in special relativity. When we define such inertial frame,  we can then use the solution of Riemann problems in a local Minkowskian frame to construct the corresponding solution in curved spacetime. The previous section derives the general relativistic hydrodynamic equations in a reference metric formulation.
For the benefit of the coming discussion, we will revert to the original formulation \cite{2000LRR.....3....2F} in this section
\begin{equation}\label{eq:discrete_form}
    \frac{1}{\sqrt{-g}} \left(\frac{\partial \sqrt{-g} \mathbf{F}^0}{\partial x^0}+\frac{\partial \sqrt{-g} \mathbf{F}^i}{\partial x^i}\right)
    =\mathbf{S}
\end{equation}
with $g = \det (g_{\mu\nu})$ satisfying $\sqrt{-g}=\alpha\sqrt{\gamma}$.
The state vector $\mathbf{F}^{0}$ and the flux vector $\mathbf{F}^{i}$ are given by 
\begin{equation}
\begin{aligned}
    \mathbf{F}^\mu 
    &= (\rho u^\mu, T^\mu_j, -n_\nu T^{\nu \mu} - \rho u^\mu) \\
    &= (\rho u^\mu, \rho h u^\mu u_j + P\delta^\mu_j, \rho h W u^\mu - P n^\mu - \rho u^\mu ),
\end{aligned}
\end{equation}
and the source term in this formulation is denoted by $\mathbf{S}$.
Since the source is irrelevant to the tetrad formulation in following discussions, we here omit the explicit form of $\mathbf{S}$.

Let us consider a single computational cell of our discrete spacetime $\Omega$,
bounded by a closed three-dimensional surface $\partial \Omega$.
We take the 3-surface $\partial \Omega$ as the standard-oriented geometric object made up of
two spacelike surfaces $\{\Sigma_{x^0}, \Sigma_{x^0+\Delta x^0}\}$ plus 
timelike surfaces $\{\Sigma_{x^i_{-}}, \Sigma_{x^i_{+}}\}$ 
that join the two temporal slices together,
where $x^i_\pm$ are the cell boundaries of $\Omega$ in $\pm x^i$ directions.
The integral form of the system (\ref{eq:discrete_form}) is 
\begin{equation}\label{eq:conservation_form}
    \int_{\Omega} \frac{1}{\sqrt{-g}} \frac{\partial \sqrt{-g} \mathbf{F}^0}{\partial x^0} d \Omega
    + \int_{\Omega} \frac{1}{\sqrt{-g}} \frac{\partial \sqrt{-g} \mathbf{F}^i}{\partial x^i} d \Omega=\int_{\Omega} \mathbf{S} d \Omega,
\end{equation}
where
\begin{align}
    d\Omega := \sqrt{-g} dx^0 \wedge dx^1 \wedge dx^2 \wedge dx^3,
\end{align}
is the volume element of cell $\Omega$.
From now we will drop the wedge symbol $\wedge$ for simplicity.
The integral form (\ref{eq:conservation_form}) can be rewritten in the following conservation form
\begin{equation}\label{eq:integral_form}
\begin{aligned}
&(\bar{\mathbf{F}}^0)_{x^0+\Delta x^0} -(\bar{\mathbf{F}}^0)_{x^0}=
 - \sum_i \left( \mathcal{F}^i_+ - \mathcal{F}^i_- \right)
 + \int_{\Omega} \mathbf{S} d \Omega 
\end{aligned}
\end{equation}
where $\left(\bar{\mathbf{F}}^0 \right)_{x^0}$ is the volume integral of $\mathbf{F}^0$ at $x^0$ given by
\begin{align}\label{eq:temporal_integral}
    \left(\bar{\mathbf{F}}^0 \right)_{x^0} &:= \int_{\Sigma_{x^0}} \sqrt{-g} \mathbf{F}^0 dx^1 dx^2 dx^3,
\end{align}
and $\mathcal{F}^i_\pm$ is the integrated spatial flux across the cell interfaces $x^i_{\pm}$ given by
\begin{equation}\label{eq:flux_integral}
    \mathcal{F}^i_\pm := \int_{\Sigma_{x^i_\pm}} \sqrt{-g} \mathbf{F}^i dx^0 \prod_{j\neq i} dx^j 
\end{equation}

\subsection{Tetrad formulation}

Instead of attempting a direct resolution of the Riemann problem within the curved spacetime, our approach entails deliberately converting the left and right states at a given interface into a local Minkowskian frame of reference. This methodology enables the utilization of developments in the realm of special relativistic Riemann problems, as proposed by \cite{1998A&A...339..638P,2006ApJ...637..296A}. 

To begin with, we define a new tetrad basis $e_{(\hat{\mu})}$ that satisfies a list of properties as shown in \cite{2016ApJS..225...22W}:
\begin{enumerate}
    \item  $e_{(\hat{\mu})}$ must be orthogonal to $e_{(\hat{\nu})}$ for all $\mu\neq \nu$. 
    \item Each  $e_{(\hat{\mu})}$ must be normalized to have an inner product of $\pm 1$ with itself, with $e_{(\hat{0})}$ being timelike and $e_{(\hat{i})}$ being spacelike.
    \item $e_{(\hat{0})}$ must be orthogonal to surfaces of constant $x^0$.
    \item The projection of $e_{(\hat{i})}$ onto to the surfaces of constant $x^0$ is orthogonal to the surface of constant $x^i$ within that submanifold. 
\end{enumerate}
Without loss of generality,
let us only consider the conversion of the volume integral $\left(\bar{\mathbf{F}}^0 \right)_{x^0}$ in Eq.~(\ref{eq:temporal_integral}) and the first spatial flux integral $\mathcal{F}^1_+$ in Eq.~(\ref{eq:integral_form}).

We define the following tetrad basis in the spherical coordinates with $x^\mu = (t, r, \theta, \phi)$
(the detailed derivation can be found in Appendix of \cite{2016ApJS..225...22W,2022PhRvD.106l4041K}) as
\begin{equation}
\begin{aligned}
&e_{(\hat{t})}{ }^\mu=n^\mu, \\
&e_{(\hat{r})}{ }^\mu=\hat{B}\left(0, \gamma^{r r}, \gamma^{r \theta}, \gamma^{r\phi}\right), \\
&e_{(\hat{\theta})}{ }^\mu=\hat{D}\left(0,0, \gamma_{\phi \phi},- \gamma_{\theta \phi}\right), \\
&e_{(\hat{\phi})}{ }^\mu=\hat{C}(0,0,0,1),
\end{aligned}
\end{equation}
where the coefficients are given by
\begin{equation}
\begin{aligned}
&\hat{B}=\frac{1}{\sqrt{\gamma^{r r}}}, \\
&\hat{C}=\frac{1}{\sqrt{\gamma_{\phi \phi}}}, \\
&\hat{D}=\frac{1}{\sqrt{\gamma_{\phi \phi}\left(\gamma_{\theta \theta} \gamma_{\phi \phi}-\gamma_{\theta \phi}^2\right)}} .
\end{aligned}
\end{equation}

The covariant components of the tetrad basis are given by $e_{(\hat{\mu})\mu} = g_{\mu \nu} e_{(\hat{\mu})}{ }^{\nu}$. Specifically
\begin{equation}
\begin{aligned}
e_{(\hat{t}) \mu} &=n_\mu, \\
e_{(\hat{r}) \mu} &=\hat{B} (\beta^r, 1, 0, 0), \\
e_{(\hat{\theta}) \mu} &=\hat{D} (\beta_\theta \gamma_{\phi \phi}-\beta_\phi \gamma_{\theta \phi}, \gamma_{r \theta} \gamma_{\phi \phi}-\gamma_{r \phi} \gamma_{\theta \phi} \\
&, \gamma_{\theta \theta} \gamma_{\phi \phi}-\gamma_{\theta \phi}^2, 0), \\
e_{(\hat{\phi}) \mu} &=\hat{C}\left(\beta_\phi, \gamma_{r \phi}, \gamma_{\theta \phi }, \gamma_{\phi \phi} \right).
\end{aligned}
\end{equation}

The transformation of vector and tensor between the tetrad frame and the original Eulerian observer frame follows
\begin{equation}
\begin{aligned}
&V_{(\hat{\mu})}=e_{(\hat{\mu})}{ }^\mu V_\mu, \\
&Q_{(\hat{\mu})(\hat{\nu})}=e_{(\hat{\mu})}{ }^\mu e_{(\hat{\nu})}{ }^\nu Q_{\mu \nu},
\end{aligned}
\end{equation}
and 
\begin{equation}
\begin{aligned}
&V_\mu=e_{(\hat{\mu}) \mu} V^{(\hat{\mu})}, \\
&Q_{\mu \nu}=e_{(\hat{\mu}) \mu} e_{(\hat{\nu}) \nu} Q^{(\hat{\mu})(\hat{\nu})}.
\end{aligned}
\end{equation}

Note that the upper and lower spatial tetrad components are the same $V^{(\hat i)} = V_{(\hat i)}$
while we have $V^{(\hat{t})} = - V_{(\hat{t})}$ for temporal component
in the local Minkowskian frame $\hat{\eta}_{(\hat \mu) (\hat\nu )} := e_{(\hat \mu )}{}^\mu e_{(\hat \nu )}{}^\nu g_{\mu\nu} = {\rm diag}(-1,1,1,1)$.

Therefore, we can define $\hat{\mathbf{F}}^{(\hat{\mu})}$ as the tetrad transformation of $\mathbf{F}^\mu$ in the form
\begin{align}
    \hat{\mathbf{F}}^{(\hat{\mu})} &=
    \begin{pmatrix}
        (\hat{F}_D)^{(\hat{\mu})} \\
        (\hat{F}_{S_{(\hat{j})}})^{(\hat{\mu})} \\
        (\hat{F}_\tau)^{(\hat{\mu})}
    \end{pmatrix}
    = e^{(\hat{\mu})}{}_\mu 
    \begin{pmatrix}
        \rho u^\mu \\
        e_{(\hat{j})}{}^\nu (\rho h u^\mu u_\nu + P \delta^\mu_\nu) \\
        \rho h W u^\mu - P n^\mu - \rho u^\mu
    \end{pmatrix},
\end{align}

Here for momentum components $(\hat{F}_{S_{(\hat{j})}})^{(\hat{\mu})}$ of $\mathbf{\hat{F}}^{(\hat{\mu})}$ we need to perform one more tetrad transformation due to its tensorial nature.
Since we only focus on the flux along $r$ direction, 
the components $\hat{\mathbf{F}}^{(\hat{t})}$ and $\hat{\mathbf{F}}^{(\hat{r})}$ are written as
\begin{subequations}
\begin{align}
    \hat{\mathbf{F}}^{(\hat{t})} &= (D, S_{(\hat{j})}, \tau), \\
    \begin{split}
    \hat{\mathbf{F}}^{(\hat{r})} &= ( D \frac{u^{(\hat{r})}}{W} , S_{(\hat{j})} \frac{u^{(\hat{r})}}{W} + P \delta^{(\hat{r})}_{(\hat{j})}
    , (\tau + P) \frac{u^{(\hat{r})}}{W} ) \\
    &= ( D \bar{v}^{(\hat{r})}, S_{(\hat{j})} \bar{v}^{(\hat{r})} + P \delta^{(\hat{r})}_{(\hat{j})}, (\tau + P) \bar{v}^{(\hat{r})} ),
    \end{split}
\end{align}    
\end{subequations}

where
\begin{align}
    u^{(\hat{t})} &= e^{(\hat{t})}{ }_\nu u^\nu = - e_{(\hat{t})t} u^t= \alpha u^t = W \\
    \begin{split}
        u_{(\hat{j})} &= e_{(\hat{j})}{}^\mu u_\mu \\
        &= ( W v^r / \sqrt{\gamma^{rr}},
        \hat{D} (u_\theta \gamma_{\phi\phi} - u_\phi \gamma_{\theta \phi}),
        \hat{C} u_\phi)
    \end{split} \\
    S_{(\hat{j})} &= \rho h W u_{(\hat{j})}, \\
    \bar{v}^{(\hat{r})} &:= \frac{u^{(\hat{r})}}{W} = v^{(\hat{r})} - \beta^{(\hat{r})} / \alpha =  \frac{v^r}{\sqrt{\gamma^{rr}}}, \label{eq:vbar}\\
    \delta^{(\hat{r})}_{(\hat{j})} &= 1 \text{ for } (\hat{r})=(\hat{j}), \text{ otherwise $0$}. 
\end{align}
The inverse transformation is given by
\begin{align}
    \mathbf{F}^{\mu} = e_{(\hat{\mu})}{}^\mu 
    \begin{pmatrix}
        (\hat{F}_D)^{(\hat{\mu})} \\
        e^{(\hat{j})}{}_j(\hat{F}_{S_{(\hat{j})}})^{(\hat{\mu})} \\
        (\hat{F}_\tau)^{(\hat{\mu})}
    \end{pmatrix},
\end{align}
which gives
\begin{align}
    \mathbf{F}^t &=\frac{1}{\alpha}
    \begin{pmatrix}
        (\hat{F}_D)^{(\hat{t})} \\
        e^{(\hat{j})}{}_j(\hat{F}_{S_{(\hat{j})}})^{(\hat{t})} \\
        (\hat{F}_\tau)^{(\hat{t})}
    \end{pmatrix}, \label{eq:Ft_transform_inv} \\
    \mathbf{F}^r &= \sqrt{\gamma^{rr}} \left[
    -\frac{\beta^{(\hat{r})}}{\alpha}
    \begin{pmatrix}
        (\hat{F}_D)^{(\hat{t})} \\
        e^{(\hat{j})}{}_j(\hat{F}_{S_{(\hat{j})}})^{(\hat{t})} \\
        (\hat{F}_\tau)^{(\hat{t})}
    \end{pmatrix}
    + 
    \begin{pmatrix}
        (\hat{F}_D)^{(\hat{r})} \\
        e^{(\hat{j})}{}_j(\hat{F}_{S_{(\hat{j})}})^{(\hat{r})} \\
        (\hat{F}_\tau)^{(\hat{r})} \label{eq:Fr_transform_inv}
    \end{pmatrix} \right].
\end{align}
In addition, we can reformulate the conservation form Eqs.~(\ref{eq:temporal_integral}) and (\ref{eq:flux_integral}) with the tetrad basis.
Note that the indexes $(t, r,\theta, \phi)$ and $(\hat{t}, \hat{r},\hat{\theta}, \hat{\phi})$
are interchangeable with $(x^0,x^1,x^2,x^3)$ and $(x^{(\hat{0})},x^{(\hat{1})},x^{(\hat{2})},x^{(\hat{3})})$ respectively.
Making use of the following invariance property
\begin{equation}
\int_{\Omega} \sqrt{-g} dt dr d\theta d\phi = \int_{\hat{\Omega}} \sqrt{-\hat{g}} d \hat{t} d \hat{r} d \hat{\theta} d \hat{\phi}
\end{equation}
and transformation rule
\begin{equation}
  dx^\mu = e_{(\hat{\nu})}{ }^{\mu} dx^{(\hat{\nu})},
\end{equation}
we can get (see also \cite{2000LRR.....3....2F})
\begin{subequations}
\begin{align}
\int_{\Sigma_{\hat{t}}} \sqrt{-\hat{g}}  d \hat{r} d \hat{\theta} d \hat{\phi}
    &=\int_{\Sigma_{t}}  \frac{1}{\alpha} \sqrt{-g}  d r d \theta d \phi, \\
\int_{\Sigma_{\hat{r}_\pm}} \sqrt{-\hat{g}} d \hat{t} d \hat{\theta} d \hat{\phi}
    &=\int_{\Sigma_{r_\pm}} \sqrt{\gamma^{rr}} \sqrt{-g} d t d \theta d \phi,
\end{align}
\end{subequations}
where $\hat{g} := \det (\hat\eta_{ij}) = -1$.
This gives the volume integral of $\mathbf{F}^t$ (\ref{eq:temporal_integral})
 and integrated spatial flux of $\mathbf{F}^r$(\ref{eq:flux_integral}) in local tetrad basis as
\begin{align} \label{eq:integral_tetrad}
\begin{split}
    \left( \bar{\mathbf{F}}^{0} \right)_{x^0} &= \int_{\Sigma_{x^0}} 
    \begin{pmatrix}
        (\hat{F}_D)^{(\hat{t})} \\
        e^{(\hat{j})}{}_j(\hat{F}_{S_{(\hat{j})}})^{(\hat{t})} \\
        (\hat{F}_\tau)^{(\hat{t})}
    \end{pmatrix} d \hat{r} d \hat{\theta} d \hat{\phi}, 
\end{split}\\
\begin{split}
    {\mathcal{F}}^{r}_\pm &= \int_{\Sigma_{r_\pm}} d \hat{r} d \hat{\theta} d \hat{\phi} \\
    &\left[
    -\frac{\beta^{(\hat{r})}}{\alpha}
    \begin{pmatrix}
        (\hat{F}_D)^{(\hat{t})} \\
        e^{(\hat{j})}{}_j(\hat{F}_{S_{(\hat{j})}})^{(\hat{t})} \\
        (\hat{F}_\tau)^{(\hat{t})}
    \end{pmatrix}
    + 
    \begin{pmatrix}
        (\hat{F}_D)^{(\hat{r})} \\
        e^{(\hat{j})}{}_j(\hat{F}_{S_{(\hat{j})}})^{(\hat{r})} \\
        (\hat{F}_\tau)^{(\hat{r})} 
    \end{pmatrix} \right],
\end{split}
\end{align}
with nonzero interface velocity 
\begin{align}
\begin{split}
    V^{(\hat{r})}_{\rm{interface}} &:= \frac{d\hat{r}}{d\hat{t}} \\
    &= \frac{\beta^{(\hat{r})}}{\alpha} = \frac{\beta^r}{\alpha\sqrt{\gamma^{rr}}},\label{eq:interface_speed}
\end{split}
\end{align}
from a nonzero drift in the direction of interest, in agreement with \cite{1998A&A...339..638P,2016ApJS..225...22W}.

With tetrad basis formulation, the procedure to obtain the numerical flux across the first spatial direction involves the following steps:
\begin{enumerate}
    \item Obtain the values of the primitive variables $(\rho, P, u^{i})$ and tetrad basis $e_{(\hat{\nu})}{}^\mu$ at $\Sigma_{x^1_\pm}$.
    \item Construct the conserved variable ${\hat{\mathbf{F}}}^{(\hat{0})}$ and flux ${\hat{\mathbf{F}}}^{(\hat{1})}$ for the left and right state in the tetrad frame.
    \item Solve the Riemann problem in the tetrad frame with a nonzero interface velocity $V^{(\hat{r})}_{\rm{interface}}$ .
    \item Once we have the updated solution of ${\hat{\mathbf{F}}}^{(\hat{0})}$ and ${\hat{\mathbf{F}}}^{(\hat{1})}$, we can obtain the numerical flux across the first spatial direction in the Eulerian observer frame according to Eq.~(\ref{eq:Fr_transform_inv}).
\end{enumerate}

\subsection{HLLC Riemann Solver in the tetrad frame}

We solve the Riemann problem in the tetrad frame by adopting a special relativity form.
We calculate the HLLC flux ${\hat{\mathbf{F}}}^{(\hat{r})}$ by solving the one-dimensional conservation law \cite{2022PhRvD.106l4041K}:

\begin{align}
& \partial_{(\hat{t})} \hat{\mathbf{U}}+\partial_{(\hat{r})} {\hat{\mathbf{F}}}^{(\hat{r})}=0, 
\end{align}
with
\begin{align}
\hat{\mathbf{U}} &:= {\hat{\mathbf{F}}^{(\hat{t})}} =
\begin{pmatrix}
    D \\
    S_{(\hat{j})} \\
    \tau
\end{pmatrix}, &
{\hat{\mathbf{F}}}^{(\hat{r})} &=
\begin{pmatrix}
    D \bar{v}^{(\hat{r})} \\
    (S_{(\hat{j})} \bar{v}^{(\hat{r})}+P \delta^{(\hat{r})}_{(\hat{j})}) \\
    (\tau + P) \bar{v}^{(\hat{r})})
\end{pmatrix}.
\end{align}

Given an initial condition at cell interface $\hat{r}_{j+\frac{1}{2}}$ described by 
\begin{equation}
\hat{\mathbf{U}}(\hat{r}, 0)= 
\begin{cases}
    \hat{\mathbf{U}}_L & \text { if } \hat{r}<\hat{r}_{j+\frac{1}{2}} \\ 
    \hat{\mathbf{U}}_R & \text { if } \hat{r}>\hat{r}_{j+\frac{1}{2}}
\end{cases},
\end{equation}
three characteristic waves and four states will be established inside the Riemann fan as
\begin{equation}\label{eq:riemann_state}
\hat{\mathbf{U}}(0, \hat{t}) = 
\begin{cases}
    \hat{\mathbf{U}}_{L} & \text { if } \quad \hat{\lambda}_{L} \geqslant V^{(\hat{r})}_{\rm{interface}} \\
    \hat{\mathbf{U}}^*_{L} & \text { if } \quad \hat{\lambda}_{L} < V^{(\hat{r})}_{\rm{interface}} \leqslant \hat{\lambda}^* \\
    \hat{\mathbf{U}}^*_{R} & \text { if } \quad \hat{\lambda}^* < V^{(\hat{r})}_{\rm{interface}} < \hat{\lambda}_{R} \\
    \hat{\mathbf{U}}_{R} & \text { if } \quad \hat{\lambda}_{R} \leqslant V^{(\hat{r})}_{\rm{interface}}
\end{cases},
\end{equation}
and the corresponding numerical flux across interface $\hat{r}_{j+\frac{1}{2}}$ is
\begin{equation}\label{eq:riemann_flux}
\begin{aligned}
& \left( {\hat{\mathbf{F}}}^{(\hat{r})}\right)_{j+\frac{1}{2}} = 
\begin{cases}
    ( {\hat{\mathbf{F}}}^{(\hat{r})})_L & \text { if } \hat{\lambda}_L\geqslant V_{\text {interface }}^{(\hat{r})} \\
    ( {\hat{\mathbf{F}}}^{(\hat{r})})^*_{L} & \text { if } \hat{\lambda}_L
        < V_{\text {interface }}^{(\hat{r})} \leqslant \hat{\lambda}^* \\
    ( {\hat{\mathbf{F}}}^{(\hat{r})})^*_{R} & \text { if } \hat{\lambda}^*<V_{\text {interface }}^{(\hat{r})} < \hat{\lambda}_R \\
    ( {\hat{\mathbf{F}}}^{(\hat{r})})_R & \text { if } \hat{\lambda}_R \leqslant V_{\text {interface }}^{(\hat{r})}
\end{cases},
\end{aligned}
\end{equation}
where $\hat{\lambda}_{L/R}$ is the characteristic speed of the left/right going nonlinear wave
and $(\hat{\mathbf{F}}^{(\hat{r})})_{L/R} := \hat{\mathbf{F}}^{(\hat{r})} (\hat{\mathbf{U}}_{L/R})$.
The intermediate state flux $( {\hat{\mathbf{F}}}^{(\hat{r})})^*_{L/R}$ may be expressed in terms of $\hat{\mathbf{U}}^*_{L/R}$ through the jump condition
\begin{equation}\label{eq:jump_condition}
( {\hat{\mathbf{F}}}^{(\hat{r})})^*_{L/R} = ( {\hat{\mathbf{F}}}^{(\hat{r})})_{L/R} + \hat{\lambda}_{L/R} ( \hat{\mathbf{U}}^*_{L/R} - \hat{\mathbf{U}}_{L/R}).
\end{equation}

Explicitly, we have the left or the right state as
\begin{subequations}
\begin{align}
  D^*\left(\hat{\lambda}-\hat{\lambda}^*\right) &=D\left(\hat{\lambda}-\bar{v}^{(\hat{r})}\right), \label{eq:1}\\
 S_{(\hat{r})}^*\left(\hat{\lambda}-\hat{\lambda}^*\right) &= S_{(\hat{r})}\left(\hat{\lambda}-\bar{v}^{(\hat{r})}\right)+P^*-P, \label{eq:2}\\
S_{(\hat{\theta})}^*\left(\hat{\lambda}-\hat{\lambda}^*\right)&=S_{(\hat{\theta})}\left(\hat{\lambda}-\bar{v}^{(\hat{r})}\right), \label{eq:3}\\
 S_{(\hat{\phi})}^*\left(\hat{\lambda}-\hat{\lambda}^*\right) &=S_{(\hat{\phi})}\left(\hat{\lambda}-\bar{v}^{(\hat{r})}\right), \label{eq:4}\\
 \tau^*\left(\hat{\lambda}-\hat{\lambda}^*\right) &=\tau\left(\hat{\lambda}-\bar{v}^{(\hat{r})}\right)+P^* \hat{\lambda}^*-P \bar{v}^{(\hat{r})}.\label{eq:5}
 \end{align}
\end{subequations}

To reduce the number of unknowns and have a well-posed problem, we assume that $S^*_{(\hat{r})}=(\tau^*+P^* + D^*) \hat{\lambda}^*$ (see \cite{2005MNRAS.364..126M}). If one defines $E:=\tau+D$ and performs the calculation of 
$({\rm \ref{eq:2}})-\hat{\lambda}^*\times[({\rm \ref{eq:1}})+({\rm \ref{eq:5}})]$, one will get the following expression, giving $\hat{\lambda}^*$ in terms of $P^*$ \cite{2005MNRAS.364..126M}:
\begin{equation}\label{eq:lambda_p}
(\hat{\lambda} E - S_{(\hat{r})} + \hat{\lambda} P^*)\hat{\lambda}^* = S_{(\hat{r})} ( \hat{\lambda}- \bar{v}_{(\hat{r})}) - P + P^*.
\end{equation}

By imposing $P^*_L= P^*_R$ across the contact discontinuity, we find the following quadratic equation for $\hat{\lambda}^*$
\begin{equation}
F^{\mathrm{HLL}}_E (\hat{\lambda}^{*})^2 - ( E^{\mathrm{HLL}} + F^{\mathrm{HLL}}_{S_{(\hat{r})}}) \hat{\lambda}^* + S_{(\hat{r})}^{\mathrm{HLL}} = 0,
\end{equation}
where 
\begin{subequations}
\begin{align}
S_{(\hat{r})}^{\mathrm{HLL}} & =\frac{\hat{\lambda}_R S_{(\hat{r})}^R-\hat{\lambda}_L S_{(\hat{r})}^L+ \hat{F}_{D, L}^{(\hat{r})}-\hat{F}_{D, R}^{(\hat{r})}}{\hat{\lambda}_R-\hat{\lambda}_L}, \\
E^{\mathrm{HLL}} & =\frac{\hat{\lambda}_R E^R-\hat{\lambda}_L E^L+\hat{F}_{E, L}^{(\hat{r})}-\hat{F}_{E, R}^{(\hat{r})}}{\hat{\lambda}_R-\hat{\lambda}_L}, \\
F_{S_{(\hat{r})}}^{\mathrm{HLL}} & =\frac{\hat{\lambda}_R \hat{F}_{S_{(\hat{r})}, L}^{(\hat{r})}-\hat{\lambda}_L \hat{F}_{S_{(\hat{r})}, R}^{(\hat{r})}+\hat{\lambda}_R \hat{\lambda}_L\left(S_{(\hat{r})}^R-S_{(\hat{r})}^L\right)}{\hat{\lambda}_R-\hat{\lambda}_L}, \\
F_{E}^{\mathrm{HLL}} & =\frac{\hat{\lambda}_R \hat{F}_{E, L}^{(\hat{r})}-\hat{\lambda}_L \hat{F}_{E, R}^{(\hat{r})}+\hat{\lambda}_R \hat{\lambda}_L\left(E^R-E^L\right)}{\hat{\lambda}_R-\hat{\lambda}_L}, \\
\hat{F}_{E, L/R}^{(\hat{r})} &:=  \hat{F}_{D, L/R}^{(\hat{r})} + \hat{F}_{\tau, L/R}^{(\hat{r})}.
\end{align}
\end{subequations}

Once we obtain the speed of the contact discontinuity $\hat{\lambda}^*$,  $P^*$ can be obtained from Eq.~(\ref{eq:lambda_p}). The conserved quantities in the intermediate states are given by
\begin{subequations}
\begin{align}
& D^*_{L / R}=\frac{D_{L / R}\left(\hat{\lambda}_{L / R}-\bar{v}_{L / R}^{(\hat{r})}\right)}{\hat{\lambda}_{L / R}-\hat{\lambda}^*}, \\
& \left(S_{(\hat{j})}\right)^*_{L /R}=\frac{1}{\hat{\lambda}_{L / R}-\hat{\lambda}^*} \\
& \times\left[\left(S_{(\hat{j})}\right)_{L / R}\left(\hat{\lambda}_{L / R}-\bar{v}_{L / R}^{(\hat{r})}\right)+\left(P^*-P_{L / R}\right) \delta^{(\hat{r})}_{(\hat{j})}\right], \\
& E^*_{L/R}=\frac{E_{L/R}\left(\hat{\lambda}_{L / R}-\bar{v}_{L / R}^{(\hat{r})}\right)+P^* \hat{\lambda}^*-P_{L / R} \bar{v}_{L / R}^{(\hat{r})}}{\hat{\lambda}_{L / R}-\hat{\lambda}^*},\\
&\tau^*_{L/R} = E^*_{L/R} - D^*_{L/R}.
\end{align}
\end{subequations}

The left and right characteristic speeds $\hat{\lambda}_{L/R}$ follow Davis's estimate \cite{2005MNRAS.364..126M}
\begin{eqnarray}
\hat{\lambda}_L &=& \min (\hat{\lambda}^-(\hat{\mathbf{U}}_L), \hat{\lambda}^-(\hat{\mathbf{U}}_R)),\\
\hat{\lambda}_R &=& \max (\hat{\lambda}^+(\hat{\mathbf{U}}_L), \hat{\lambda}^+(\hat{\mathbf{U}}_R)),
\end{eqnarray}
with
\begin{equation}\label{eq:lambda_hat}
\begin{aligned}
\hat{\lambda}^{\pm} & =\frac{1}{1-v^2 c_s^2}\left[\bar{v}^{(\hat{r})}\left(1-c_s^2\right)\right. \\
& \left.\pm c_s \sqrt{\left(1-v^2\right)\left(1-v^2 c_s^2-\left(1-c_s^2\right)\left(\bar{v}^{(\hat{r})}\right)^2\right)}\right],
\end{aligned}
\end{equation}
where $v^2 =\bar{v}^{(\hat{i})} \bar{v}_{(\hat{i})}$ and $c_s$ is the speed of sound
\begin{equation}
c_s^2 =\frac{1}{h}\left[\left.\frac{\partial P}{\partial \rho}\right|_{\varepsilon}+\left.\frac{P}{\rho^2} \frac{\partial P}{\partial \varepsilon}\right|_\rho\right].
\end{equation}

Equivalent expressions for the $(\theta,\phi)$ directions can be easily obtained.
In the Eulerian observer frame, the minimum and maximum characteristic speeds $\lambda^{\pm}$ are given by \cite{1999astro.ph.11034I,Banyuls_1997,2000LRR.....3....2F}:

\begin{equation}\label{eq:lambda}
\begin{aligned}
  & \lambda^{\pm}= -\beta^r + \frac{\alpha}{1-v^2 c_s^2}\left [v^r\left(1-c_s^2\right)\right. \\
&\left. \pm  c_s \sqrt{\left(1-v^2\right)\left[\gamma^{r r}\left(1-v^2 c_s^2\right)-v^r v^r\left(1-c_s^2\right)\right]} \right] 
\end{aligned}.
\end{equation}

Making use of Eq.~(\ref{eq:vbar}) and Eq.~(\ref{eq:lambda_hat}), we can get the following relation:
\begin{equation}
\lambda^{\pm} = -\beta^r  + \alpha \sqrt{\gamma^{rr}} \hat{\lambda}^{\pm}. \label{eq:contact_speed}
\end{equation}

\added[id=1]{Note that Eq.~(\ref{eq:lambda_hat}) and Eq.~(\ref{eq:lambda}) are derived in the special relativistic and general relativistic setting, respectively. Equation~(\ref{eq:contact_speed}) establishes their relationship with the tetrad method consistently.}

\added[id=1]{In addition, we implement the HLLE Riemann solver \cite{doi10.11371025002,1988SJNA...25..294E} for comparison. We adopt the same tetrad formulation. The HLLE Riemann solver is constructed by assuming an average intermediate state between the fastest and slowest waves in the tetrad frame. The two characteristic waves and three states inside the Riemann fan become:}
\begin{equation}\label{eq:hlle_riemann_state}
\hat{\mathbf{U}}(0, \hat{t}) = 
\begin{cases}
    \hat{\mathbf{U}}_{L} & \text { if } \quad \hat{\lambda}_{L} \geqslant V^{(\hat{r})}_{\rm{interface}} \\
    \hat{\mathbf{U}}^* & \text { if } \quad \hat{\lambda}_{L} < V^{(\hat{r})}_{\rm{interface}} < \hat{\lambda}_{R} \\
    \hat{\mathbf{U}}_{R} & \text { if } \quad \hat{\lambda}_{R} \leqslant V^{(\hat{r})}_{\rm{interface}}.
\end{cases}
\end{equation}

\added[id=1]{The corresponding numerical flux across interface $\hat{r}_{j+\frac{1}{2}}$ is:}

\begin{equation}
\begin{aligned}
& \left( {\hat{\mathbf{F}}}^{(\hat{r})}\right)_{j+\frac{1}{2}} = 
\begin{cases}
    ( {\hat{\mathbf{F}}}^{(\hat{r})})_L & \text { if } \hat{\lambda}_L\geqslant V_{\text {interface }}^{(\hat{r})} \\
    ( {\hat{\mathbf{F}}}^{(\hat{r})})^* & \text { if } \hat{\lambda}_L<V_{\text {interface }}^{(\hat{r})} < \hat{\lambda}_R \\
    ( {\hat{\mathbf{F}}}^{(\hat{r})})_R & \text { if } \hat{\lambda}_R \leqslant V_{\text {interface }}^{(\hat{r})}
\end{cases},
\end{aligned}
\end{equation}

\added[id=1]{where $\hat{\mathbf{U}}^*$ and $(\hat{\mathbf{F}}^{(\hat{r})})^*$ are the intermediate state and flux. They can be derived from the jump condition [see Eq. \ref{eq:jump_condition}] as:}
\begin{equation}
  \hat{\mathbf{U}}^* = \frac{\hat{\lambda}_R \hat{\mathbf{U}}_R - \hat{\lambda}_L \hat{\mathbf{U}}_L -  ( {\hat{\mathbf{F}}}^{(\hat{r})})_R +  ( {\hat{\mathbf{F}}}^{(\hat{r})})_L} {\hat{\lambda}_R  - \hat{\lambda}_L},
\end{equation}
and
\begin{equation}
 ( {\hat{\mathbf{F}}}^{(\hat{r})})^* = \frac{ \hat{\lambda}_R ( {\hat{\mathbf{F}}}^{(\hat{r})})_L - \hat{\lambda}_L  ( {\hat{\mathbf{F}}}^{(\hat{r})})_R + \hat{\lambda}_R \hat{\lambda}_L (\hat{\mathbf{U}}_R - \hat{\mathbf{U}}_L ) }{ \hat{\lambda}_R - \hat{\lambda}_L}.
\end{equation}

\subsection{HLLC Riemann Solver for the Moving-Mesh GR}

For the moving mesh in the simulation domain, naturally, we need to solve the Riemann problem on the moving interface with its own coordinate velocity $\mathbf{V}:= \frac{d \mathbf{x}}{d t} = (V^r, V^\theta, V^\phi)$. Let us denote the corresponding four-velocity as $u_{\rm{gridface}}^{\mu}$. In general, when we consider the $3+1$ spacetime foliation $\Sigma_t$, we define a unit normal vector as $n^\mu$, and this unit normal vector corresponds by deﬁnition to the four-velocity of
the Eulerian observer \cite{10.1093/acprof:oso/9780199205677.001.0001}. When we define the fluid's four-velocity as $u^\mu$, the velocity of the fluid with respect to the Eulerian observer ($v^\mu=(0,v^i)$) has the following relation:
\begin{equation}
u^{\mu}=W (n^\mu + v^\mu),
\end{equation}
where $W= (1-g_{\mu\nu}v^{\mu}v^{\nu})^{-1/2}$ is the Lorentz factor of the fluid with respect to the Eulerian observer.
When we move from a given hypersurface to the next following the normal direction, the change in the spatial coordinates is given as \cite{10.1093/acprof:oso/9780199205677.001.0001}:
\begin{equation}
  x_{t+d t}^i=x_t^i-\beta^i dt,
\end{equation}
$\mathbf{\beta}^\mu = (0,\mathbf{\beta}^i)$ being the shift vector. Then 
$v^i$ is related to the coordinate velocity $\mathbf{V}$ by $v^i = \frac{1}{\alpha}(V^i + \beta^i)$.
In our case, only the cell interface orthogonal to the radial direction can move with a coordinate velocity denoted as $V^r$.
Then the four-velocity of our radially-moving interface is 
\begin{eqnarray}
u_{\rm{gridface}}^{\mu} &=& W \left( \frac{1}{\alpha}, \frac{V^r}{\alpha}, 0, 0 \right),\\
v^i &=& \left( \frac{V^r+\beta^r}{\alpha}, \frac{\beta^\theta}{\alpha}, \frac{\beta^\phi}{\alpha} \right)
\end{eqnarray}

In the above, we illustrate the explicit definition of different velocities for clarity. For our moving-mesh code, the grid moves radially, the integral of the radial flux at a short time interval $dt$ becomes

\begin{equation}
\int_{\Sigma_{r}} \sqrt{-g} (\mathbf{F}^r - V^r \mathbf{F}^t)  d t d \theta d \phi
\end{equation}

With

\begin{equation}
    V^r = \frac{dx^r}{dt} = \frac{e_{(\hat{\mu})}{ }^r dx^{(\hat{\mu})}}{e_{(\hat{\nu})}{ }^t dx^{(\hat{\nu})}} = -\beta^r + \alpha \sqrt{\gamma^{rr}} V^{(\hat{r})},
\end{equation}

Note that the above velocity equation relates to Eq.~(\ref{eq:interface_speed}) and Eq.~(\ref{eq:contact_speed}). From Eqs. (\ref{eq:Ft_transform_inv}), and (\ref{eq:Fr_transform_inv}), we have:

\begin{equation}
\mathbf{F}^r - V^r \mathbf{F}^t
= \sqrt{\gamma^{rr}}(\hat{\mathbf{F}}^{(\hat{r})} - V^{(\hat{r})} \hat{\mathbf{F}}^{(\hat{t})})
\end{equation}

Compared with the tetrad formulation for the static mesh, we replace the interface velocity $V^{(\hat{r})}_{\rm{interface}} = \frac{\beta^r}{\alpha \sqrt{\gamma^{rr}}}$ by $V^{(\hat{r})}_{\rm{interface}} =  V^{(\hat{r})} = \frac{(\beta^r + V^r)}{\alpha \sqrt{\gamma^{rr}}} $ to incorporate the effect of the moving interface into the flux integral. 

In principle, the coordinate velocity for the moving interface can be set freely. At each instantaneous time, on the cell interface, the three characteristic waves and four states inside the Riemann fan depends only on the values of the primitive variables on the left and right sides of the interface. The interface velocity will influence which state the numerical flux across the interface will be selected [see Eqs.~(\ref{eq:riemann_state}], and (\ref{eq:riemann_flux}) ). Based on this flexibility, we choose the contact discontinuity velocity as the interface velocity:
  \begin{equation}
    V^{(\hat{r})}_{\rm{interface}} = \hat{\lambda}^*, V^r_{\rm{interface}} = -\beta^r + \alpha \sqrt{\gamma^{rr}}\hat{\lambda}^*.
  \end{equation}

\added[id=1]{We find this choice performs well for the simulation of ultrarelativistic jets.}

  For the derivation of the tetrad formulation and HLLC Riemann solver, we express every metric and fluid variable in the coordinate basis. For the implementation, we utilize those variables in the orthonormal basis instead. \added[id=1]{For example, in the tetrad basis calculation, we will use $\gamma_{\{i\}\{j\}}=\psi^4 \bar{\gamma}_{\{i\}\{j\}} = \psi^4(\delta_{\{i\}\{j\}} + \epsilon_{\{i\}\{j\}})$ instead of $\gamma_{ij}$ itself. In this way, the geometric factors will not directly appear in the tetrad basis calculation. The derivation itself remains the same because of} the invariance of spacetime interval under coordinate transformation
\begin{equation}
  ds^2 = g_{\mu\nu}dx^{\mu}dx^{\nu} = g_{\{\mu\}\{\nu\}} dx^{\{\mu\}}dx^{\{\nu\}}.
\end{equation}

Making use of this invariance principle, we can handle the moving mesh in another way. First, boost the coordinate basis into the comoving coordinate basis of the interface:
\begin{equation}
    dx^{\langle \mu \rangle} = \frac{\partial x^{\langle \mu \rangle}}{\partial x^{\nu}} dx^{\nu} = \Lambda_{\nu}{ }^{\langle \mu \rangle}  dx^{\nu}
\end{equation}
Second, boost the primitive velocities into the comoving coordinate basis:
\begin{equation}
  v^{\langle \mu \rangle} = \Lambda_{\nu}{ }^{\langle \mu \rangle} v^{\nu}
\end{equation}
Third, making use of the invariance, calculate the corresponding metric $g_{\langle \mu \rangle \langle \nu \rangle}$ components:
\begin{equation}
  ds^2 = g_{\langle\mu\rangle \langle\nu\rangle }dx^{\langle\mu\rangle}dx^{\langle\nu\rangle} = g_{\mu \nu} dx^{\mu}dx^{\nu}
\end{equation}
Once we have the new lapse, shift and spatial metric in the comoving frame, we can derive the tetrad basis in the comoving coordinate basis, and solve the HLLC Riemann problem accordingly. We lay out this approach for readers' interest as well as for a more complete discussion.

\section{Numerical Techniques}\label{sec:techniques}

\subsection{Implementation of equations}

For the numerical implementation, we discretize the volume averages of Eq.~(\ref{eq:form_eq}). Using divergence theorem, the discretized version of equation \ref{eq:form_eq} in the cell $(i, j)$ can be expressed as (Since our code is 2.5D, we will ignore the discretization in the $\phi$ direction.) \cite{2021MNRAS.508.2279C}:
\begin{equation}
    \begin{aligned}
&\frac{\mathrm{d}}{\mathrm{d} t}\langle\boldsymbol{q}\rangle_{i, j}= \frac{1}{\Delta V_{i, j}} \\
& \times\left\{\left[\left.\left(\langle\boldsymbol{f}\rangle^{1} \Delta \mathcal{A}^{1}\right)\right|_{i+1 / 2, j}-\left.\left(\langle\boldsymbol{f}\rangle^{1} \Delta \mathcal{A}^{1}\right)\right|_{i-1 / 2, j}\right]\right.\\
&+\left[\left.\left(\langle\boldsymbol{f}\rangle^{2} \Delta \mathcal{A}^{2}\right)\right|_{i, j+1 / 2}-\left.\left(\langle\boldsymbol{f}\rangle^{2} \Delta \mathcal{A}^{2}\right)\right|_{i, j-1 / 2}\right]\} \\
&+\langle\boldsymbol{s}\rangle_{i, j}
\end{aligned}
\end{equation}

where the cell volume and volume average are defined as 
\begin{equation}
    \begin{aligned}
&\Delta V \equiv \int_{\text {cell }} \sqrt{\hat{\gamma}} \mathrm{d} x^{1} \mathrm{~d} x^{2} \mathrm{~d} x^{3}, \\
&\langle\bullet\rangle \equiv \frac{1}{\Delta V} \int_{\text {cell }} \bullet \sqrt{\hat{\gamma}} \mathrm{d} x^{1} \mathrm{~d} x^{2} \mathrm{~d} x^{3},
\end{aligned}
\end{equation}
while the surface area and surface average is deﬁned as
\begin{equation}
    \begin{aligned}
&\Delta \mathcal{A}^{i} \equiv \int_{\text {surface }} \sqrt{\hat{\gamma}} \mathrm{d} x^{j, j \neq i} \\
&\langle\bullet\rangle^{i} \equiv \frac{1}{\Delta A^{i}} \int_{\text {surface }} \bullet^{i} \sqrt{\hat{\gamma}} \mathrm{d} x^{j, j \neq i}
\end{aligned}.
\end{equation}
Note that when we perform the volumn average $\langle\bullet\rangle$ or surface average $\langle\bullet\rangle^{i}$, we could strip out the geometric factor from the tensorial expressions in coordinate basis and integrate them together with the volume factor $\sqrt{\hat{\gamma}}$. In this way, the tensorial variables in orthonormal basis like $S_{\{\theta\}}$ become truly independent of the underlining geometry. For example, in the spherical coordinates, the volume average for the conserved momentum $q_{S_\theta}$ will be calculated as 
\begin{align}
\begin{split}
\langle q_{S_\theta} \rangle_{i,j} \Delta V
&= \int_{\text {cell }} \mathcal{M} S_{\{\theta\}}  r^3 \sin \theta d r d \theta d \phi \\
&= \left( \mathcal{M} S_{\{\theta\}} \right)_{i,j} \int_{\text {cell }}  r^3 \sin \theta d r d \theta d \phi.
\end{split}
\end{align}
For our moving-mesh scheme, the cells in the radial direction will continuously merge and divide. When we perform the above integral, the variables in orthonormal basis like $\mathcal{M}S_{\{\theta\}}$ will be better conserved. As an example, if we assume $\mathcal{M} S_{\{\theta\}}$ is constant across ${\rm{cell}}_{(i,j)}$ and ${\rm{cell}}_{(i+1,j)}$, when we merge these two cells, it gives a combined conserved momentum as
\begin{align}
\begin{split}
\langle q_{S_\theta} \rangle_{i_{\rm{new}},j} &\Delta V_{i_{\rm{new}},j} = 
    \langle q_{S_\theta} \rangle_{i,j} \Delta V_{i,j} + \langle q_{S_\theta} \rangle_{i+1,j} \Delta V_{i+1,j} \\
&= \left( \mathcal{M} S_{\{\theta\}} \right)_{i_{\rm{new}},j} \int_{{\rm{cell}}_{(i_{\rm{new}},j)}}  r^3 \sin \theta d r d \theta d \phi,
\end{split}
\end{align}
where ${\rm cell}_{(i_{\rm{new}},j)}$ is the combined cells of ${\rm cell}_{(i,j)}$ and ${\rm cell}_{(i+1,j)}$ with $\Delta V_{i_{\rm{new}},j} = \Delta V_{i,j} + \Delta V_{i+1,j}$. From the combined momentum, we can recover the variable $\mathcal{M} S_{\{\theta\}}$ accurately. 

\added[id=1]{In code implementation, the contribution of the source term to the conserved variables inside a cell is defined as $\langle \boldsymbol{s}\rangle_{i, j} \Delta V_{i,j}$. We perform volume integral on the singular factors such as $1/r$ and $\rm{cot}(\theta)$ that appear in the source term [see Eq.~(\ref{eq:complete_eqs})].  Explicitly, the integral of $1/r$ factor gives $(r_{+}^2-r_{-}^2)/2$ while the integral of $\rm{cot}(\theta)$ leads to $(\rm{sin}(\theta_{+}) -\rm{sin}(\theta_{-}))$. This practice turns out to reduce numerical error for the source term calculation near singular points.}

Finally, to work out the cell volume, cell surface, we make the following definition 
\begin{equation}
    r_{\pm}=r \pm \frac{1}{2} \Delta r, \quad \theta_{\pm}=\theta \pm \frac{1}{2} \Delta \theta, 
\end{equation}
and calculate the area and volume as
\begin{equation}
    \begin{aligned}
&\left.\Delta \mathcal{A}_{r}\right|_{r_{\pm}}= 
    2\pi \left(r \pm \frac{\Delta r}{2}\right)^{2}\left(2 \sin \theta \sin \left(\frac{\Delta \theta}{2}\right)\right), \\
&\left.\Delta \mathcal{A}_{\theta}\right|_{\theta_{\pm}}= 
    2\pi \left(r^{2}+\frac{1}{12}(\Delta r)^{2}\right) \left(\Delta r\right)\left(\sin \left(\theta \pm \frac{\Delta \theta}{2}\right)\right), \\
&\Delta V= 2 \pi
    \left(r^{2}+\frac{1}{12}(\Delta r)^{2}\right) \left(\Delta r\right)\left(2 \sin \theta \sin \left(\frac{\Delta \theta}{2}\right)\right).
\end{aligned}
\end{equation}

\subsection{Recovery of primitive variables}\label{sec:prim_rec}

There are many possible ways to make the conversion between conserved variables and primitive variables (e.g., \cite{2006ApJ...641..626N}). Our current research focuses on relativistic jets propagating in an ambient medium. We need to deal with large variations of density and pressure in the jet simulations. The following cons-to-prim method proves to be robust for such a task. We use $\rho, P,\ Wv^{\{i\}}= u^{\{i\}} + W \beta^{\{i\}}/\alpha $ as our primitive variables where $Wv^{\{i\}}$ is the projected fluid velocity in orthonormal basis. For the equation of state (EOS), we only consider the case of a single-component perfect gas for now. In this case, the specific enthalpy $h \equiv 1 + \varepsilon + P/\rho$ is a function of a temperaturelike variable $\Theta = P/\rho$ (see \cite{1958PhT....11l..56S}).
In the literature, the most widely used EOS is the ideal gas EOS:$P=(\Gamma-1)\rho\varepsilon$, where $P$ is the gas pressure,  $\varepsilon$ is the specific internal energy density. Which can be expressed as:
\begin{equation}
  h(\Theta) = 1+ \frac{\Gamma}{\Gamma-1}\Theta,
\end{equation}
where $\Gamma$ is the adiabatic index. The ideal gas EOS has been applied to the gas of either subrelativistic temperature with $\Gamma=5/3$ or ultrarelativistic temperature with $\Gamma=4/3$. For our simulations of relativistic outflow propagating in a cold ambient medium,  a variable equivalent adiabatic index $\Gamma_{\rm{eq}} = (h-1)/(h-1-\Theta)$ is desirable to account for transitions between the nonrelativistic and the relativistic temperature regime. There have been efforts to find EOSs that better describe the thermal dynamics of relativistic gas. \citeauthor{1958PhT....11l..56S} derives the correct EOS for the single-component perfect gas in a relativistic regime using modiﬁed Bessel functions. \citeauthor{2005ApJS..160..199M} proposes an approximate EOS (denoted as TM EOS) that is consistent with the Taub’s
inequality \cite{1948PhRv...74..328T}:
\begin{equation}
  (h-\Theta)(h-4\Theta) \geq 1
\end{equation}
for all temperatures. It differs by less than $4\%$ from the theoretical value given in \cite{1958PhT....11l..56S}. \citeauthor{2006ApJS..166..410R} proposes a new EOS (RC EOS), which better fits the theoretical value. Let us write the expression of the specific enthalpy for the RC EOS:

\begin{equation}\label{eq:RC_EOS}
h(\Theta) = 2 \frac{6\Theta^2 + 4\Theta +1 }{3\Theta + 2}.
\end{equation}

Following the definition of the general form of polytropic index $n$ and the general form of sound speed $c_s$:

\begin{equation}
n = \rho \frac{\partial h}{\partial p} - 1, c_s^2 = -\frac{\rho}{nh} \frac{\partial h}{\partial \rho},
\end{equation}
their values can be calculated for RC EOS as:
\begin{eqnarray}\nonumber 
  n&=&3 \frac{9 \Theta^2+12 \Theta+2}{(3 \Theta+2)^2}, \\
   c_s^2&=&\frac{\Theta(3 \Theta+2)\left(18 \Theta^2+24 \Theta+5\right)}{3\left(6 \Theta^2+4 \Theta+1\right)\left(9 \Theta^2+12 \Theta+2\right)}.
\end{eqnarray}
For both TM and RC, we have correctly $c_s^2 \rightarrow 5\Theta/3$ in the nonrelativistic temperature limit and $c_s^2\rightarrow 1/3$ in the ultrarelativistic temperature limit \cite{2006ApJS..166..410R}.

We can use these expressions to convert the conservative variables into primitive ones with a standard Newton–Raphson method (NRM) \cite{2012ApJ...746..122D}, using $\Theta$ as our independent variable. We will use the (known) values of the conservative variables. 
\begin{equation}
D = \rho W, S_{\{i\}} = \rho h W^2 v_{\{i\}}, \tau = \rho h W^2 - P -D
\end{equation}
First, by squaring the momentum equation, we get 
\begin{equation}\label{eq:lorentz}
W^2 = 1 + \frac{S^2}{D^2 h^2}, S^2 = \gamma^{\{i\}\{j\}}S_{\{i\}} S_{\{j\}}
\end{equation}
with $h=h(\Theta)$ given by the EOS. 
Using the relation $p=D\Theta /W$, we get the energy density (excluding rest mass), $\tau = DhW - D\Theta/W - D$. 
We can then derive the following identity \cite{2012ApJ...746..122D} :

\begin{equation}
f(\Theta) = h(\Theta)W(\Theta) - \frac{\Theta}{W(\Theta)} - 1 - \frac{\tau}{D} = 0.
\end{equation}
Together with Eq.~(\ref{eq:lorentz}), the derivative $df/d\Theta$ has the form:
\begin{equation}
\frac{df(\Theta)}{d\Theta} = \frac{h^\prime}{W} \left( 1- \frac{\Theta}{h}\frac{W^2-1}{W^2}\right) - \frac{1}{W}
\end{equation}
where the relation $W^\prime = - h^\prime (W^2-1)/(hW)$ has been used (derived from Eq.~(\ref{eq:lorentz}), see also \cite{2012ApJ...746..122D}). 

The derivative $dh/d\Theta$ depends on the particular EOS used. We adopt the RC EOS [see Eq.~(\ref{eq:RC_EOS})] for the simulations of relativistic jets and the ideal gas EOS for the remaining numerical tests. 

\subsection{Reconstruction}\label{sec:reconstruction}
We reconstruct the primitive variable (denoted with $Q$ ) to the left and right sides of each cell with the total variation diminishing (TVD) method described in \cite{2014JCoPh.270..784M}:
\begin{equation}
  Q_i^{\pm} = \left <Q \right >_i + \overline{\Delta Q_i} \frac{\xi_{i\pm \frac{1}{2}} - \bar{\xi}_i}{\Delta \xi_i}
\end{equation}

where $\Delta \xi_i = \xi_{i+1/2} -\xi_{i-1/2}$ is the cell width, and $\bar{\xi}$ is the cell center. And $\overline{\Delta Q_i}$ is a slope-limited gradient function written in terms of a nonlinear limiter function $\varphi(v)$:
\begin{eqnarray}
  \overline{\Delta Q}_i&=&\Delta Q_i^F \varphi(v) \quad \text { where } v=\frac{\Delta Q_i^B}{\Delta Q_i^F} \\
  \Delta Q_i^F&=&\Delta \xi_i\left(\frac{\langle Q\rangle_{i+1}-\langle Q\rangle_i}{\bar{\xi}_{i+1}-\bar{\xi}_i}\right) \\
  \Delta Q_i^B&=&\Delta \xi_i\left(\frac{\langle Q\rangle_i-\langle Q\rangle_{i-1}}{\bar{\xi}_i-\bar{\xi}_{i-1}}\right)  
\end{eqnarray}

We adopt the same modified monotonized central (MC) limiter in \cite{2014JCoPh.270..784M}. 
\begin{eqnarray}
  &\varphi^{M C}(v)&=\max \left[0, \min \left(\frac{1+v}{2}, c_i^F, c_i^B v\right)\right] \\
  &\text { where }& c_i^F=\frac{\bar{\xi}_{i+1}-\bar{\xi}_i}{\xi_{i+\frac{1}{2}}-\bar{\xi}_i} ; c_i^B=\frac{\bar{\xi}_i-\bar{\xi}_{i-1}}{\bar{\xi}_i-\xi_{i-\frac{1}{2}}}
\end{eqnarray}

To reconstruct the left and right state of the cell $i$, the above stop limiter utilizes the cell average values of $\left<Q\right>_{i-1}$,$\left<Q\right>_{i}$, and $\left<Q\right>_{i+1}$, defined at the cell center $\bar{\xi}_{i-1},\bar{\xi}_{i},\bar{\xi}_{i+1}$. This algorithm takes into account nonuniform spacing. The cell center position $\bar{\xi}_i$ can be taken as the volume-averaged cell center (``centroids of volume'') or arithmetic-mean cell center. In this study, we adopt the arithmetic-mean cell center for our simulations.

\subsection{Treatment of numerical conditions}
\added[id=1]{
  Robust numerical simulations require the treatment of several numerical conditions. One of them is the Courant-Friedrich-Levy (CFL) condition \cite{1967IBMJ...11..215C}, which limits the time step size in explicit numerical methods. The simulation domain of the \textit{JET} code allocates cells at the same temporal level. A global time step will be used to evolve simulation time. To find the global time step, we first calculate the time step $\delta t$ of individual cells in the domain according to:}
  \begin{eqnarray}
    \delta t &=&\textit{CFL}\cdot \rm{min}( \delta t^r, \delta t^{\theta}) \\
    \delta t^r &=& \frac{\Delta r}{\rm{max}(|(\lambda^{r})^{+} - V^r_{\rm{cell}}|,|(\lambda^{r})^{-} - V^r_{\rm{cell}}|)}  \label{eq:dt_r} \\
    \delta t^{\theta} &=& \frac{ r\Delta \theta}{ \rm{max}(|(\lambda^{\theta})^{+}|, |(\lambda^{\theta})^{-}|)}
  \end{eqnarray}
  \added[id=1]{where $\textit{CFL}$ is the CFL number. Its value has been set to 0.4 for simulations performed in this study. $(\lambda^r)^{\pm}$ and $(\lambda^{\theta})^{\pm}$ are again the minimum and maximum characteristic speeds for the cell in the radial and polar direction, respectively.  $V^{r}_{\rm{cell}}$ is the cell's radial velocity which approximates the cell's upper interface velocity. We then pick the smallest time step as the global one. The subtraction of the cell's radial velocity in Eq.~(\ref{eq:dt_r}) leads to a much larger time step, making the long-term simulation of relativistic jets computationally efficient.}
    \added[id=1]{Another numerical condition that needs to be taken care of is the boundary condition. For our cell-centered grid structure in spherical polar coordinates, we follow the boundary treatment described in \cite{2013PhRvD..87d4026B,2020PhRvD.101j4007M}. We first allocate two layers of ghost zones for each of the four boundaries (two in the radial direction, and two in the polar direction), and then fill the boundary ghost zones at the radial origin, and at the $\theta$ boundary with values copied from the corresponding points in the interior of the grid, accounting for appropriate parity factors. For the outer boundary in the radial direction, we adopt the Dirichlet boundary condition and use the initial data routine to set their ghost zone values.}
  
\subsection{The adjusted moving-mesh scheme}
\added[id=1]{Since the initial development of the \textit{JET} code \cite{2013ApJ...775...87D}, the moving-mesh scheme has kept being updated to improve the accuracy and efficiency of relativistic jet simulations. The adjusted moving-mesh scheme in this study contains the following rules: inside the simulation domain, the radial interface of a grid cell will move at local contact discontinuity velocity of the flow. Each radial track moves independently. The inner and outer radial boundaries of the domain can also move. At each time step, the longest and shortest cell in each radial track will be marked for refinement or derefinement according to the maximum or minimum aspect ratio of grid cell ($a:=\Delta r/ r\Delta \theta$) allowed in the simulation (see \cite{2013ApJ...775...87D} for more information). In ultrarelativistic jet simulations, we find the domain cells can squeeze into an ultrathin shell with the cell's aspect ratio reaching 1/100 or even smaller. In order to resolve the relativistic thin shell, only cells with length $\Delta r < r/(8\Gamma^2)$ will be marked for derefinement. In addition, we deﬁne an approximate second derivative of a ﬂuid variable as a measurement of error $E_i$ to mark the region of interest. At each time step, the cell along each radial track with the maximum measurement of error will be marked for refinement if its aspect ratio is larger than twice the minimum aspect ratio and its measurement error $E_i>0.9$. The cells with $E_i < 0.002$ will be considered for derefinement. The cell to be dereﬁned is the one that has the smallest time step (see \cite{2019ApJ...880..135X}). To reduce load imbalance of CPUs, the number of grids in each radial track will be balanced dynamically during the simulation.}

\section{Fixed-Mesh Numerical Simulations}\label{sec:fixed_mesh_test}

\subsection{Bondi accretion in maximally sliced trumpet coordinates}\label{sec:bondi_accretion}

\begin{figure}[!ht]
  \centering
    \includegraphics[clip,width=\columnwidth]
                    {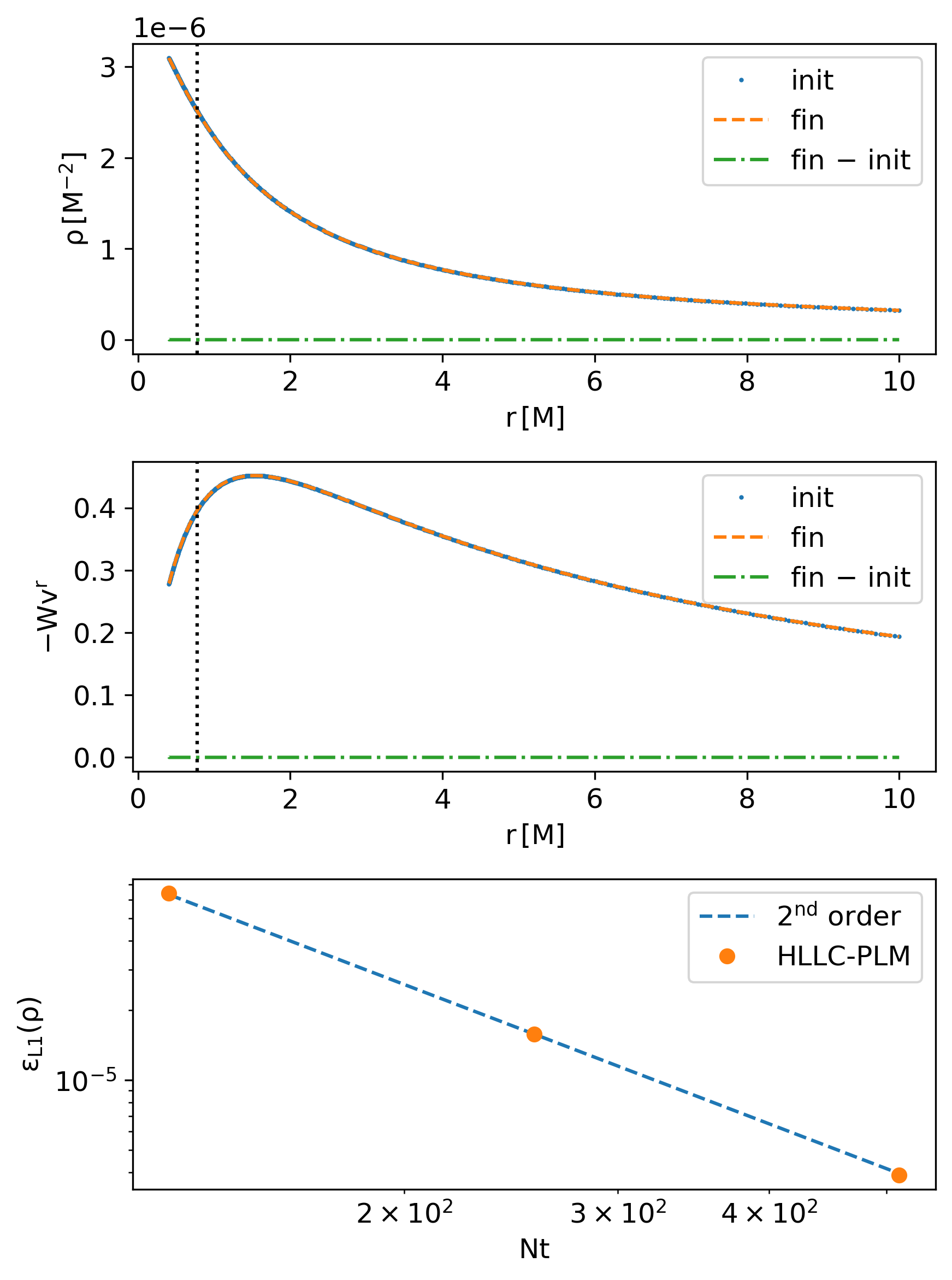}
  \caption{Top: radial rest-mass density profile of the Bondi flow for the medium resolution $\texttt{Nt}=256$. The mass accretion rate is $\dot{M}_{\rm{bondi}}=10^{-4}\,M$. The blue dotted line indicates the initial analytical solution at $T=0\,M$, while the yellow dashed line denotes the numerical solution at $T=50\,M$. The green dash-dotted line shows their difference. The black dotted vertical line indicates the radial location of the event horizon of the black hole at $r=0.78\,M$ in the maximally sliced trumpet coordinate. Middle: the radial fluid velocity $-W v^r$ profile. $W$ is the Lorentz factor, and $v^r$ is the fluid velocity with respect to the Eulerian observer.  Bottom: the L1-norm of error for the rest-mass density as a function of the azimuthal grid number $\texttt{Nt}$. The dash line indicates second-order convergence.} \label{fig:bondi}
\end{figure}

We first consider spherically symmetric, radial fluid accretion onto a nonrotating black hole (ingoing Bondi flow) \cite{1952MNRAS.112..195B,1972Ap&SS..15..153M}. Following previous work (e.g., \cite{2017CQGra..34c5007M,2022PhRvD.106l4041K}), we perform simulations of Bondi flow in maximally sliced trumpet coordinates \cite{PhysRevD.75.067502,2008PhRvD..78f4020H}. 
The transformation between Schwarzschild coordinate and maximally slicing trumpet coordinate is illustrated as a reference in Appendix \ref{appendix:bondi}. We set the fluid parameter according to Table 1 of \cite{2017CQGra..34c5007M}: the accretion rate $\dot{M}_{\rm{bondi}}=10^{-4} M$, the adiabatic index $\Gamma=4/3$, and the critical radius $R_s=10 M$ where M is the mass of the central black hole. For simplicity, $M$ is set to 1 in the simulation.

The simulation domain is in an axisymmetric spherical coordinate, spanning the region $r\in[0.4\,M,10\,M],\ \theta \in[0,\pi/2]$. 
We employ logarithmic grid spacing in the radial direction with a cell's aspect ratio $a$ set to one (i.e. $a:=\Delta r/ r\Delta \theta = 1$). 
The finest cell, located closest to the inner boundary, has a spacing $\Delta r = r_{\rm{min}}\, (\theta_{\rm{max}}-\theta_{\rm{min}})/(\texttt{Nt})$,
where \texttt{Nt} is the number of cells in the azimuthal direction. To maintain the unity aspect ratio of the cell, the number of cells in the radial direction \texttt{Nr} is calculated as
\begin{equation}
  \texttt{Nr}=\frac{\log(r_{\rm{max}}/r_{\rm{min}})}{\log(1+(\theta_{\rm{max}}-\theta_{\rm{min}})/ \texttt{Nt})}.
\end{equation}
We conduct simulations with three different resolutions: low resolution with $\texttt{Nt}=128$, medium resolution with $\texttt{Nt} = 256$, and high resolution with $\texttt{Nt} = 512$. For the benefit of convergence test, we set the number of grids in the radial direction $\texttt{Nr}=2\,\texttt{Nt}$. In this case, the cell's aspect ratio will deviate from one slightly.  

In Fig.~\ref{fig:bondi}, we show the radial profiles of the fluid rest-mass density (top) and the fluid velocity (middle) at time $T=0\,M\,(\rm{init})$ and $T=50\,M\,(\rm{fin})$ for the medium resolution simulation. The profile of the Bondi flow has been maintained throughout the simulations. 
In the bottom panel, we plot the L1-norm of error for the rest-mass density. The L1-norm of error is defined as \cite{2016ApJS..225...22W}
\begin{equation}
  \epsilon_{\rm L1}(\rho) = \frac{\int |\rho_{\rm{fin}}-\rho_{\rm{init}}| \sqrt{-g} drd\theta d\phi}{\int |\rho_{\rm{init}}| \sqrt{-g} drd\theta d\phi}.
\end{equation}

The Bondi simulations demonstrate second-order convergence for the L1-norm of error with respect to the resolution. The code adopts the second-order RK2 time integrator and the second-order piecewise linear reconstruction method (PLM), described in Sec. \ref{sec:reconstruction}. The presented convergence result is as expected and agrees with previous studies (see e.g. \cite{2016ApJS..225...22W,2022PhRvD.106l4041K}). For the implementation of a higher-order reconstruction scheme for our unstructured grid in spherical geometry, like the piecewise parabolic method (PPM) \cite{1984JCoPh..54..174C}, weighted essentially nonoscillatory (WENO) \cite{LIU1994200,1996JCoPh.126..202J,1997JCoPh.136...83S,2007MNRAS.379..469T}, or the monotonicity preserving scheme (MP5) \cite{1997JCoPh.136...83S},  we will refer to future work.

\subsection{Tolmann-Oppenheimer-Volkoff star}

The next numerical test we consider is the Tolman–Oppenheimer–Volkoff (TOV) star with the structure of a spherically symmetric body of isotropic material in equilibrium \cite{PhysRev.55.364,PhysRev.55.374}.

\begin{table}[htb]
\centering
\caption{Parameter values for the initial profile of TOV star. \label{table:tov}}
\begin{tabular}{ p{2cm}|p{2cm}|p{2cm}|p{2cm} }
 \hline
 \hline
  Radius [km] & Gravitational mass $[M_{\odot}]$  & Baryon mass $[M_{\odot}]$ & $\rho_c\ [\rm{g/cm^3}]$  \\
 \hline
   12 &  1.40  & 1.51     & $7.92 \times 10^{14}$ \\
   \hline
   \hline
\end{tabular}  
\end{table}

We conduct two TOV star tests based on\cite{2020MNRAS.496..206C}: the stationary case and the one with pressure depletion.
The initial profile for the TOV star has a central rest-mass density $\rho_c(0)=1.28\times 10^{-3}$.
We adopt the polytropic EOS $ P(\rho)=K\rho^\Gamma$, with $(K, \Gamma) = (100, 2)$ for the initial data. As for the evolution, we adopt the ideal gas law. Additional parameters for the initial profile can be found in
Table \ref{table:tov} in the cgs unit.

\begin{figure}[!ht]
  \centering
    \includegraphics[clip,width=\columnwidth]
                    {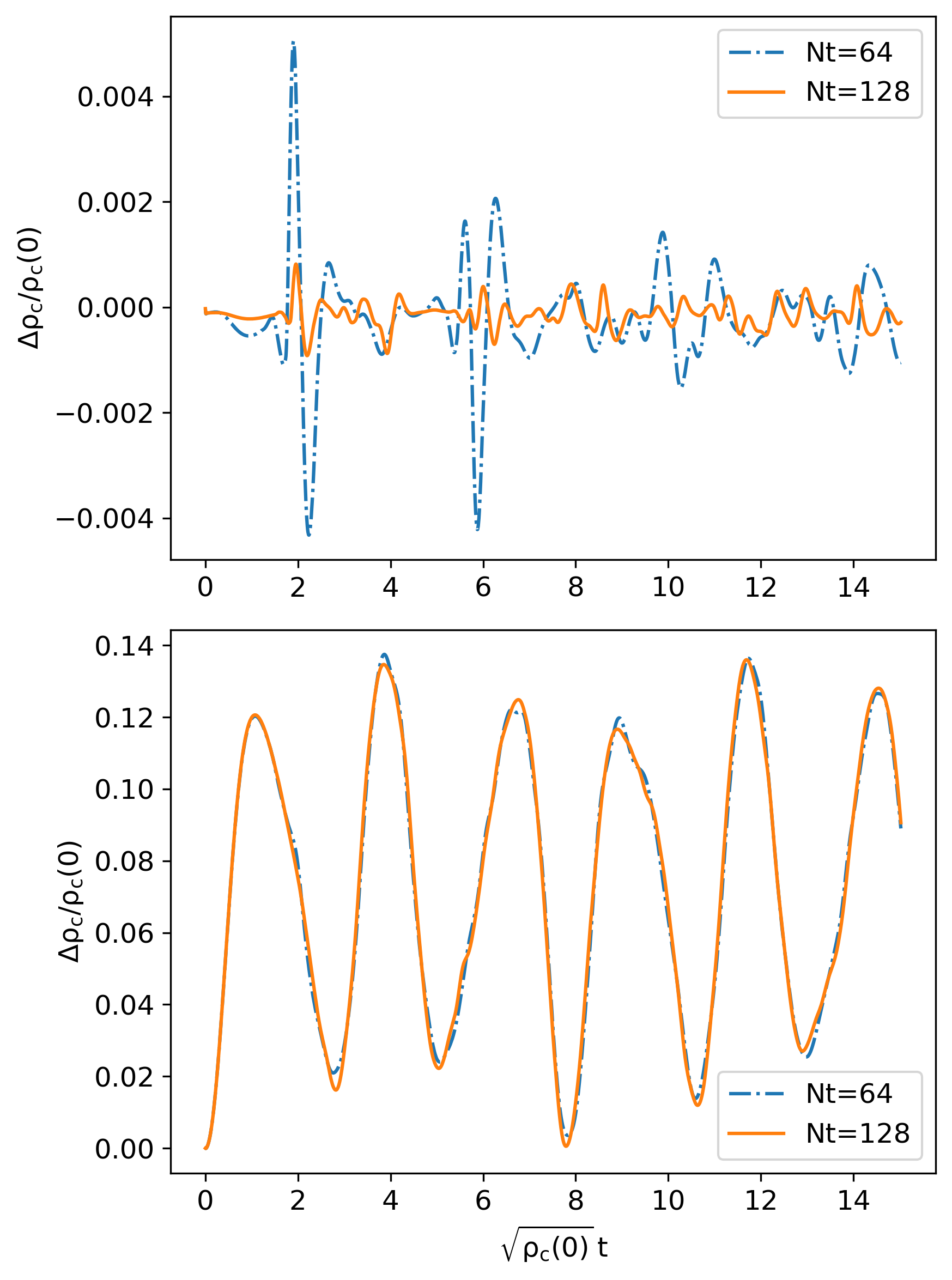}
  \caption{Normalized variation for the central maximum density as a function of dynamical time for the TOV star at two resolutions ($\texttt{Nt}=64$ and $\texttt{Nt}=128$). Top: the time evolution for the stationary case. 
Bottom: the time evolution for the pressure-depleted star whose equilibrium pressure has been reduced by ten percent globally.\label{fig:tov}}
\end{figure}

In Fig.~\ref{fig:tov}, we plot the central maximum density variation as a function of dynamical time ($\sqrt{\rho_c(0)}\,t$) for both cases. 
For the stationary case, we find the central maximum density varies within 0.5\% for 14 dynamical times for the $\texttt{Nt}=64$ simulation, confirming the stability of the star. When we increase the resolution to $\texttt{Nt}=128$, the result gets better. For the pressure depletion simulation, we reduce the TOV initial pressure profile by ten percent. The star falls out of equilibrium and undergoes radial oscillations. We conduct simulations with two different resolutions ($\texttt{Nt}=64$ and $128$) and find consistent oscillation pattern, as shown in the bottom panel of Fig.~\ref{fig:tov}. The result is equivalent to the test result in \cite{2020MNRAS.496..206C}.

\subsection{Fishbone-Moncrief torus around a Schwarzschild black hole}

Our next test concerns a stationary, axisymmetric, isentropic torus around a Schwarzschild black hole \cite{1976ApJ...207..962F}. We consider a particular instance of the Fishbone-Moncrief solution where the spin of the black hole is set to zero.

We generate the initial data in the Schwarzschild coordinate with its radius denoted by $R$. However, we will evolve the system 
in the isotropic coordinate of the Schwarzschild metric with its radius denoted by $r$ (see Appendix \ref{appendix:bondi}). The initial profile generator follows the implementation in \cite{2003ApJ...589..444G,2017ComAC...4....1P,2018PhRvD..97f4036R}. Table \ref{table:torus} shows the key variable values for the torus. For the ambient atmosphere, we set $\rho =\rho_{\rm{min}} (R/R_g)^{-3/2}, P=P_{\rm{min}}(R/R_g)^{-5/2}$, where $\rho_{\rm{min}}=10^{-8},P_{\rm{min}}=10^{-10}$, $R_g= GM/c^2$ is the black hole gravitational radius and $M$ is the black hole mass. 

\begin{table}[htb]
\centering
\caption{Parameters for the stationary torus around a black hole
\footnote{$M$ is the mass of the central Schwarzschild black hole. $R_{in}$ is the location of the inner edge of the torus. 
$R_{\rm{max}}$ is the pressure maximum location of the torus in Schwarzschild coordinates. $\rho_{\rm{max}}$ is the peak density in the torus, $\Phi_{R_{\rm{max}}}$ is the angular velocity at $R_{\rm{max}}$.(calculated from the simulation with $\texttt{Nt}=512$ resolution). $l=u^tu_{\phi}$ is the constant specific angular momentum. $\kappa$ and $\Gamma$ define the EOS of the initial torus $P=\kappa \rho^\Gamma$. Unless specified, the presented variable value follows the $G=c=M_{\odot}=1$ unit convention.}
}\label{table:torus}
\begin{tabular}{ p{2cm}|p{2cm}|p{2cm}|p{2cm}|p{2cm}}
 \hline
 \hline
 $M ~[M_\odot]$ & $R_{\rm{in}}$&  $R_{\rm{max}}$  & $\Phi_{R_{\rm{max}}}$ \\
 \hline
  1 & 6 & 12  & $2.63\times 10^{-2}$  \\
 \hline
 \hline
 $\rho_{\rm{max}}$ & $l$ & $\kappa$ & $\Gamma$  \\
 \hline
 $6.86\times 10^{-3}$ & 4.62 & $1\times 10^{-3}$  & 4/3 \\
 \hline
\end{tabular}
\end{table}

\begin{figure}[!htb]
  \centering
    \includegraphics[clip,width=\columnwidth]
                    {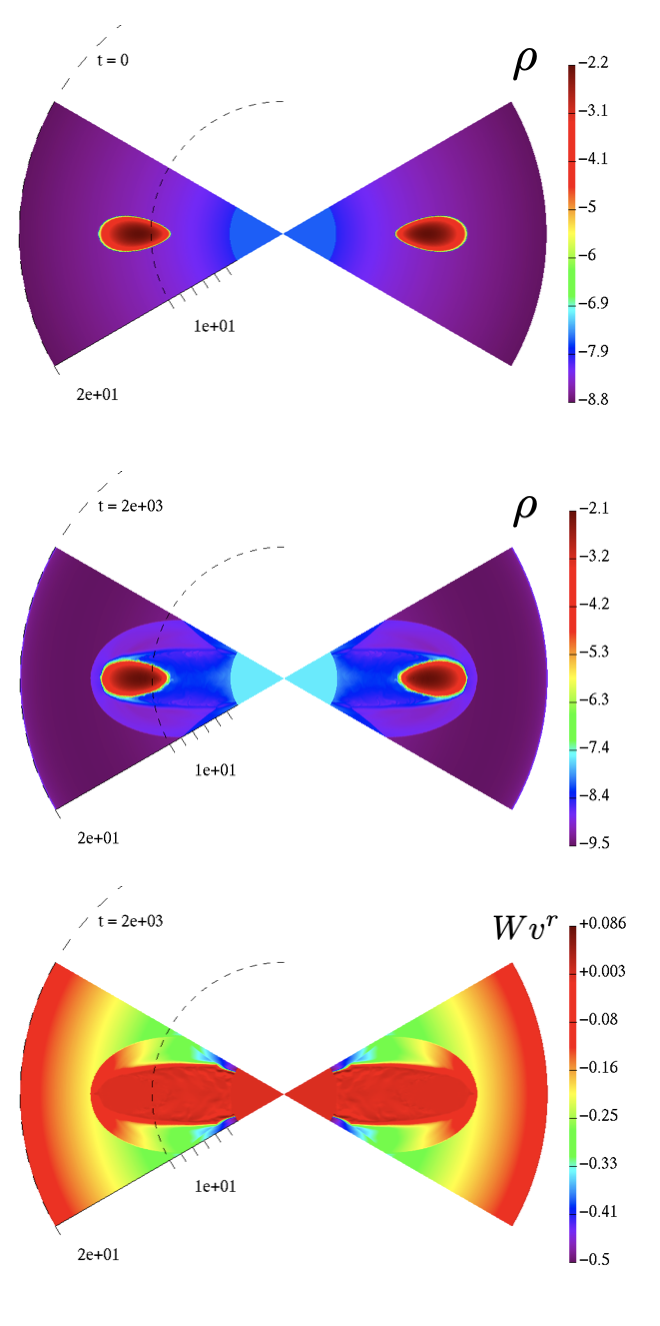}
  \caption{Contour plot of a stationary torus around a static black hole. The simulation adopts an axisymmetric spherical domain. The azimuthal angle extends from $\pi/3$ to $2 \pi/3$. The radial grid extends from $r_{\rm{min}} = 4$ to $r_{\rm{max}}=20$. The top panel shows the initial profile of the logarithmic density. The middle panel represents the density profile at the end of the simulation $t=2000$. The bottom panel represents the contour plot of the radial velocity $W v^r$. } \label{fig:torus}
\end{figure}

For the simulation, we employ an ideal gas EOS: $P = (\Gamma-1) \rho \varepsilon$, with $\Gamma=4/3$. In the azimuthal direction, the simulation domain extends from $\theta_{\rm{min}}=\pi/3$ to $\theta_{\rm{max}}= 2\pi/3$. In the radial direction, the grid covers the region from $r_{\rm{min}}=4 $ to $r_{\rm{max}} = 20$. At the location of maximum pressure $r=10.98\,(R=12)$, the orbital period of the torus is around $238.9$. We set the final time of the simulation to be $2000$, roughly eight orbits. We conduct two simulations with grid resolution $\texttt{Nt} = 256$ and $\texttt{Nt}=512$, and find consistent results.

Figure~\ref{fig:torus} illustrates the contour plots of the black hole-torus system at the beginning (top panel) and at the end of the simulation (middle and bottom panels), taken from the $\texttt{Nt}=512$ simulation for better visual effect. The top panel shows the initial contour plot for the logarithmic density $\rho$. Comparing these two contour snapshots, we first find that throughout the simulation, the torus maintains its density structure. We check that the maximum rest-mass density always keeps
the original value within 4\% during the simulation, and its radial position varies within 2\%. Because the torus stays close to the black hole, the ambient gas falls into the black hole and blows the torus surface in the infalling process. A bow shock appears in front of the torus and a trailing tail fills in the inner region between the torus and the central black hole. The falling gas slows down when it crosses the bow shock as can be seen from the velocity contour plot. The stability of the torus structure near the black hole showcases the code's robustness in the handling of fluid rotation under strong gravity.

\subsection{Rayleigh-Taylor instability for a modified Bondi flow}

\begin{figure*}[htb]
  \centering
  \includegraphics[width=\textwidth]{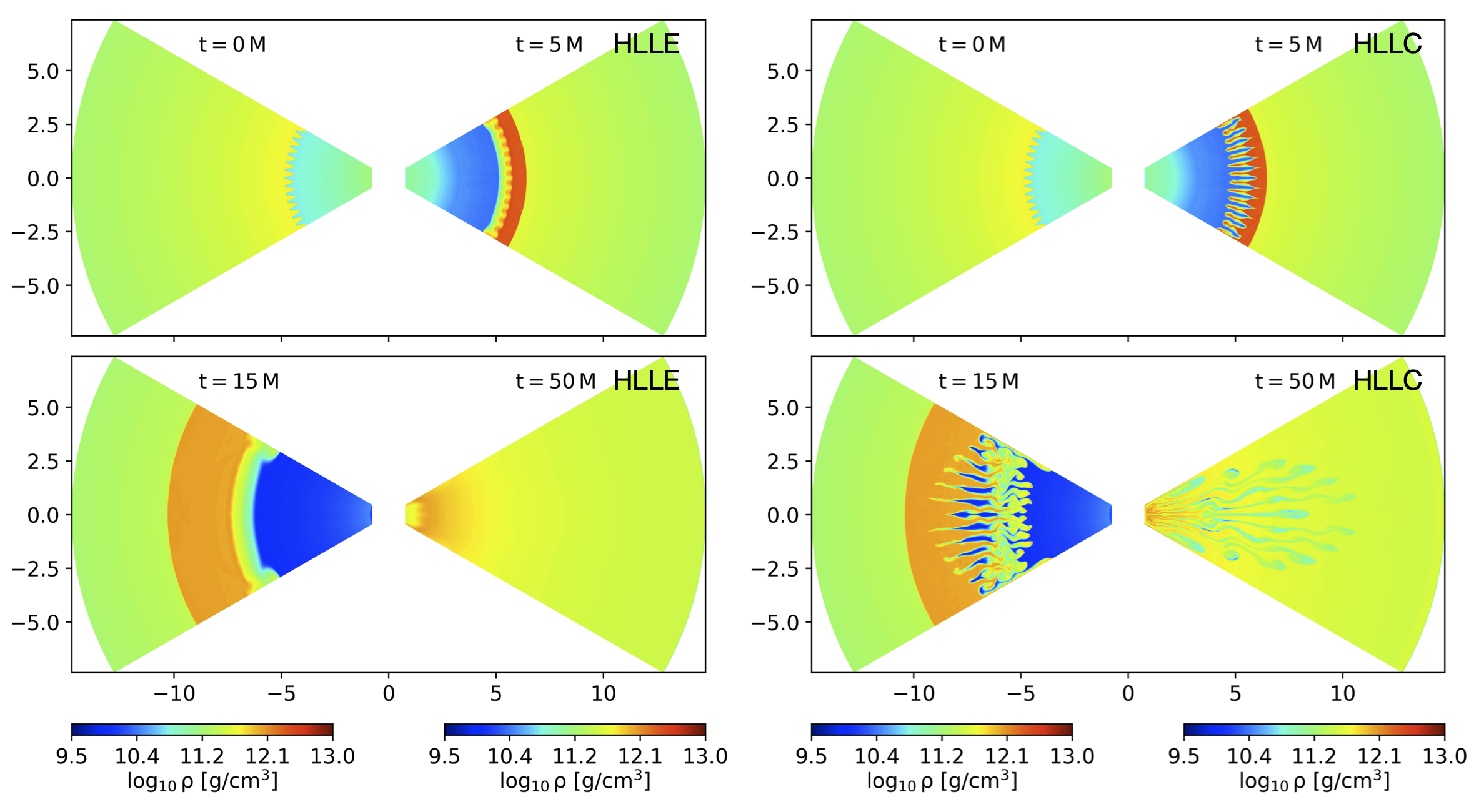}
  \caption{Time snapshots for the density contour of a hot low-density bubble embedded inside a cold Bondi flow. \added[id=1]{The left panel shows the simulation results with the HLLE Riemann solver while the right panel presents the comparative results from the simulation with the HLLC Riemann solver.} The simulation video for the case with HLLC Riemann solver is available from Youtube at \cite{videoRT}. } \label{fig:bondi_rt}
\end{figure*}

Previous work \cite{2013ApJ...775...87D} with the original \texttt{JET} code has captured the detailed nonlinear features of Rayleigh-Taylor instability in a relativistic fireball. It uses the HLLC Riemann solver described in \cite{2011ApJS..197...15D}. To test our general relativistic HLLC Riemann solver, we modify the Bondi flow to induce Rayleigh-Taylor instability under strong gravity. The setup is similar to a Str\"omgren sphere around the central black hole--the low-density hot gas is surrounded by a high-density gas with gravitational acceleration \cite{2014MNRAS.437.2856P}. Within a radius of $r < 3(1+0.1\times (1+\cos (80\theta))/2.0)$, the density and pressure of the Bondi flow have been modified as $\rho=0.1\rho_{\rm{bondi}},P=50P_{\rm{bondi}},v^r=0$. $\rho_{\rm{bondi}}$ and $P_{\rm{bondi}}$ are taken from the Bondi profile in Sec. \ref{sec:bondi_accretion}. This setup creates a hot low-density bubble inside the Bondi flow with a curly interface. As the hot low-density gas pushes against the heavier Bondi flow, Rayleigh-Taylor instability (or sometimes referred to as Richtmyer Meshkov instability in this case) develops. We perform this simulation with an azimuthal resolution of $\texttt{Nt=512}$, covering the azimuthal angle from $\pi/3$ to $2\pi/3$. Figure~\ref{fig:bondi_rt} shows its time evolution. Initially, the hot gas pushes outward and compresses the incoming Bondi flow into higher density as shown at $t=5\,\rm{M}$. Instability fingers develop and evolve inside the low-density region. Nonlinear features of the instability continuously evolve at $t=15\,\rm{M}$. Later on, due to the attraction of the central black hole, the turbulent gas flows into the black hole. The implemented HLLC Riemann solver is able to capture the detailed structure of the instability in the strong field regime. It performs better than the HLLE Riemann solver. 

\section{Moving-Mesh Numerical Simulations}\label{sec:moving_mesh_test}
\subsection{Spherical shock tube test}

\begin{figure}[htb]
  \centering
  \includegraphics[width=\columnwidth]{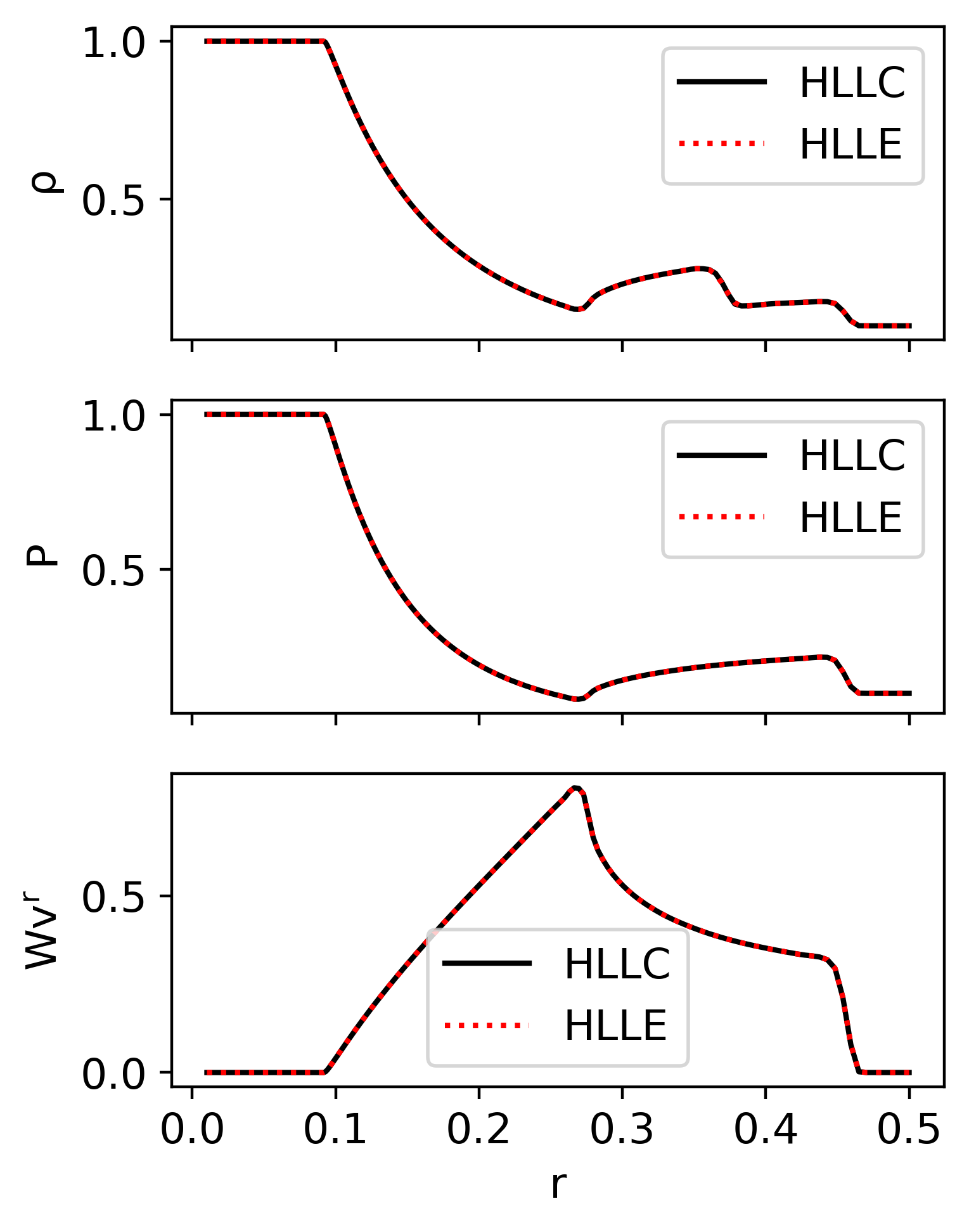}
  \caption{\added[id=1]{Profiles for the spherical shock tube test at $t=0.3$ in fixed-mesh simulations with the HLLC (solid line) and HLLE (dotted line) Riemann solver.} } \label{fig:grsod_hllc_hlle}
\end{figure}

\begin{figure*}[hbtp]
  \centering
  \subfloat[]{
    \includegraphics[width=0.5\textwidth]{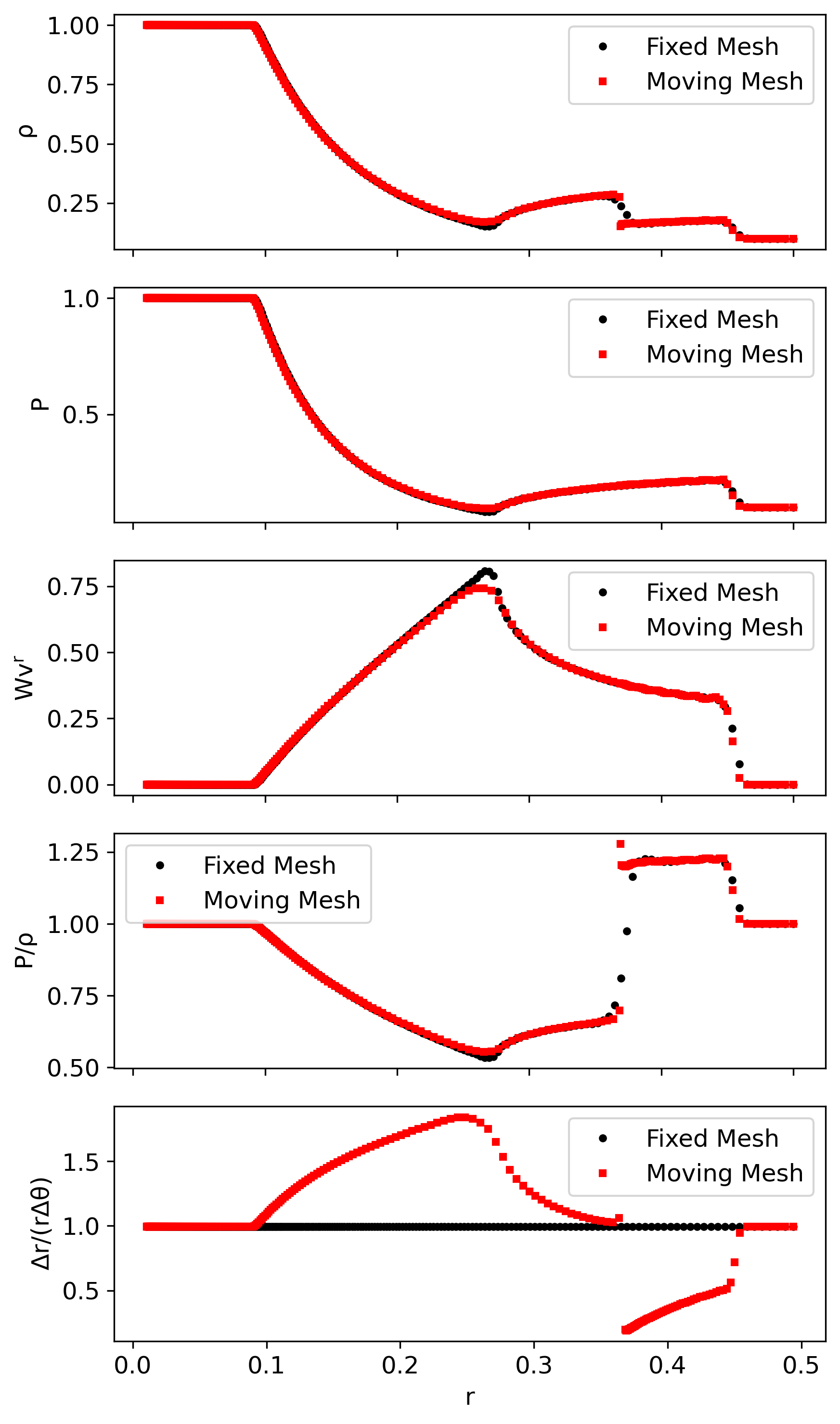}
    \label{fig:sod}
  }
  \subfloat[]{
    \includegraphics[width=0.5\textwidth]{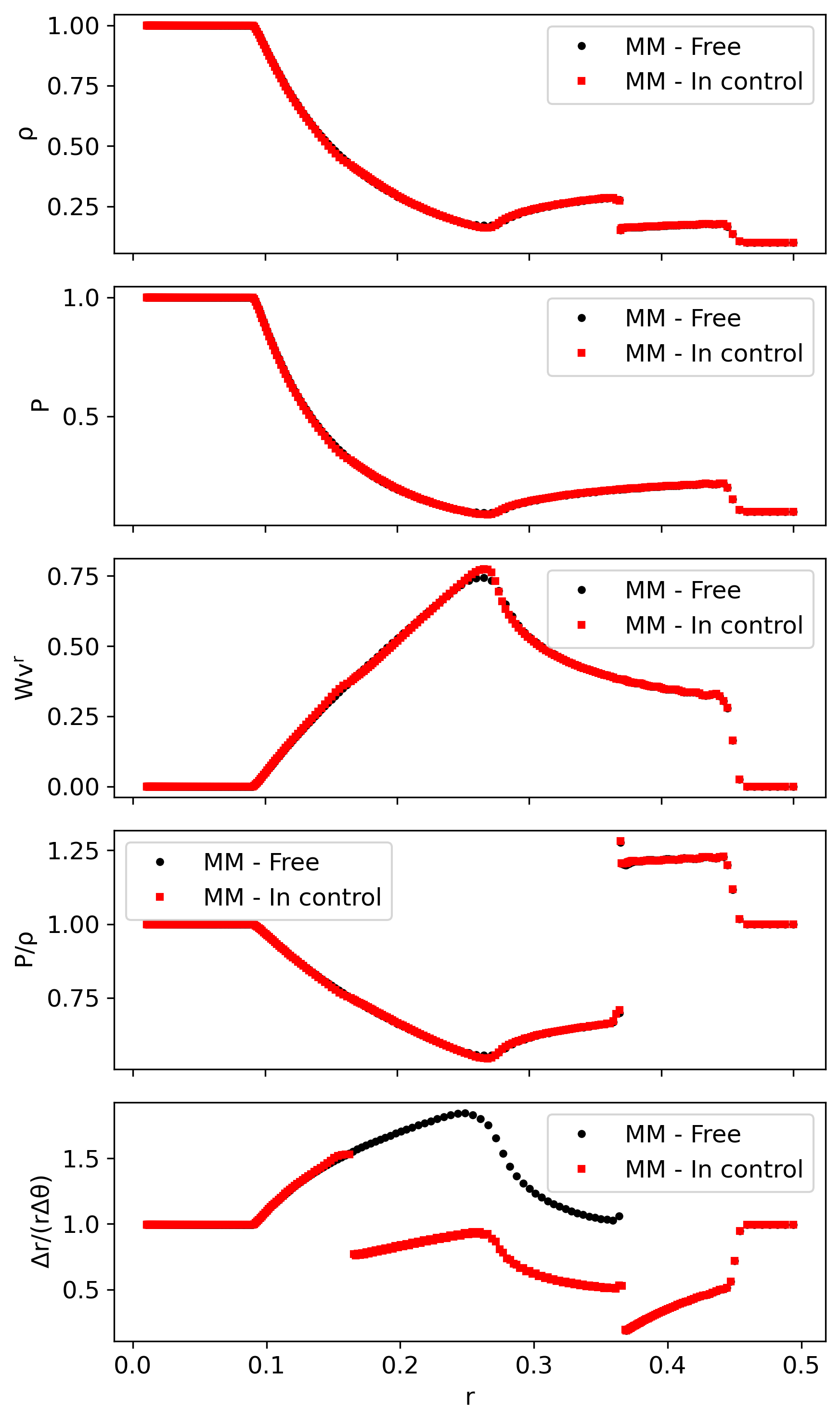}
    \label{fig:sod_mm_comparison}
  }
  \caption{ (a) Shock tube test in spherical coordinate for the fixed-mesh (black-dot) and moving-mesh (MM) (red-square) simulations. The profiles are presented at $t=0.3$. Initial discontinuity locates at $x=0.25$. The inner (outer) state of the shock tube is $\rho=1,P=1,v=0$ ($\rho=0.1,P=0.1,v=0$). The EOS follows the ideal gas law with $\Gamma=4/3$. (b)  Shock tube test in spherical coordinate for the moving-mesh simulations with different control schemes for the cell's aspect ratio. The black-dot line represents the simulation where the cell's maximum aspect ratio has not been set. The red-square line shows the simulation where the maximum aspect ratio of the cell has been set to 1.5.}
\end{figure*}

One advantage of our moving-mesh code is that the cell face is able to move with the contact velocity of the flow in the radial direction. It has been shown that the contact discontinuity is much better preserved when employing HLLC on the moving mesh (see Fig. 7 of \cite{2011ApJS..197...15D}). What is more, the flow naturally adjusts the cell width in the radial direction. Combined with robust refinement and derefinement schemes, the simulation domain will be able to resolve the region of interest \cite{2019ApJ...880..135X}. To test the accuracy of the moving-mesh scheme, we conduct the identical spherical shock tube test as shown in \cite{2022MNRAS.510.1315A}: within the radius of 0.25 ($r<0.25$), the density $\rho$ and pressure $P$ is set to 1. Outside of this region, the value of density and pressure is 0.1. We adopt the Minkowskian frame for the test. Since the tetrad formulation for the HLLC Riemann solver also works for the Minkowskian metric, we do not take any additional steps for the special relativistic simulations.

In azimuthal direction, the simulation domain extends from 0 to $\pi/2$ with $\texttt{Nt}=128$. In the radial direction, the grid covers the region from $r_{\rm{min}} = 0.01$ to $r_{\rm{max}} = 0.5$. We adopt logarithmic spacing in the radial direction and set the initial cell's aspect ratio to one. \added[id=1]{We first conduct the spherical shock tube test with different Riemann solvers in fixed-mesh simulation. Both the HLLE and HLLC Riemann solver handle the test well and give almost the same results (as shown in Fig.~\ref{fig:grsod_hllc_hlle}).}
In Fig.~\ref{fig:sod}, we compare the end profile for simulations with the fixed mesh setup and the moving mesh setup. 
The density plot exhibits a sharp transition at the contact discontinuity for the moving mesh and a relatively smooth one for the fixed mesh. Following the compression of the fluid in the shocked region, the cells squeeze between the contact discontinuity and the forward shock. The $P/\rho$ plot reveals a jump at the contact discontinuity. We find this appears in the moving-mesh simulation here as well as in the literature \cite{2011ApJS..197...15D,2022MNRAS.510.1315A}.  It may come from the physical squeezing of the fluid as the grid moves together \added[id=1]{with the flow} in the moving-mesh simulation or the TVD reconstruction scheme requires some adjustification for the moving mesh.

\begin{figure}[htb]
  \centering
  \includegraphics[width=\columnwidth]{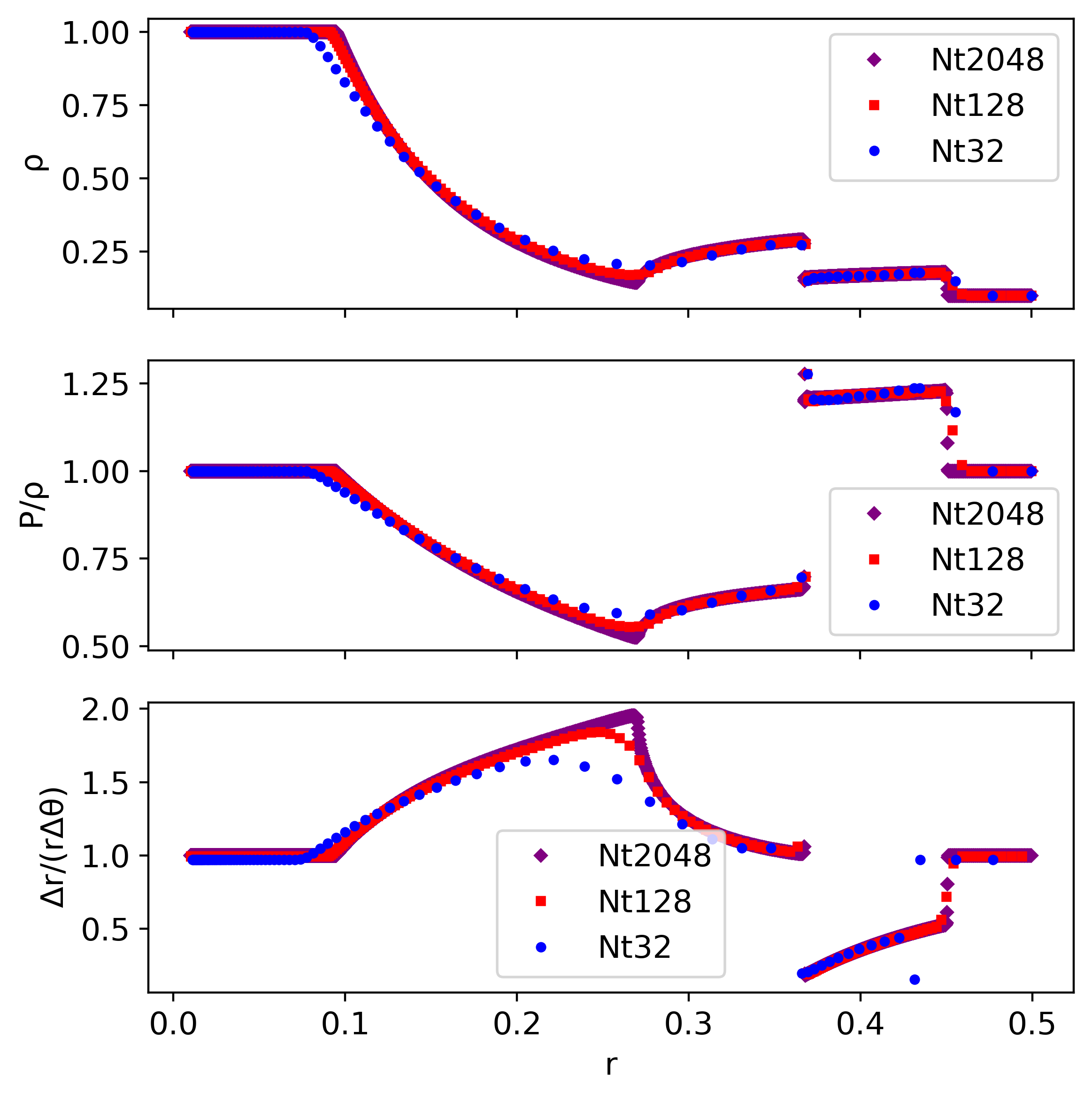}
  \caption{\added[id=1]{Profiles for the spherical shock tube test at $t=0.3$ in the moving-mesh simulations with different grid resolution. The purple diamond shows results from the simulation with resolution $\texttt{Nt}=2048$ while the red square and blue circle represent results from the $\texttt{Nt}=128$ simulation and the $\texttt{Nt}=32$ simulation, respectively.} }\label{fig:grsod_move_res}
\end{figure}

\added[id=1]{To investigate the jump's dependence on numerical resolution, we have performed additional moving-mesh simulations with different numerical resolution: the $\texttt{Nt}=32, \texttt{Nt}=128$ and $\texttt{Nt}=2048$ resolution. Results from these three simulations (see Fig.~\ref{fig:grsod_move_res}) demonstrate that the jump feature persists and its magnitude is invariant under different numerical resolution. We also perform additional fixed-mesh simulations with higher resolution and find no presence of the jump feature in these simulations.} Since the jump's magnitude does not increase with time \added[id=1]{and spatial resolution}, \replaced[id=1]{considering its minimal impact on the fluid dynamics in moving-mesh simulations}{ it has minimal impact on the fluid dynamics}, we will leave this numerical phenomena to the research community for now.

In the rarefaction region, the cells get elongated, leading to an aspect ratio larger than one. Because of the increase in the aspect ratio (i.e. the reduction of radial resolution), we find the peak of the velocity profile for the moving-mesh simulation becomes less sharp compared to the fixed mesh simulation.


However, since we have full control over the grid refinement, we can specify the maximum aspect ratio in the simulation. We conduct another moving-mesh sod-tube simulation which sets the maximum aspect ratio to 1.5. When the elongated cell reaches such a threshold, it will split into two cells. To show the effect of such a refinement scheme on the sod-tube simulation,  we compare the profiles for the moving-mesh simulation with or without maximum aspect ratio control in Fig.~\ref{fig:sod_mm_comparison}. 
With the maximum aspect ratio control, the resolution in the region where the cell's aspect ratio gets to the threshold value increases. The peak of the velocity profile becomes sharper compared to the peak for the moving-mesh simulation without aspect ratio control. Overall, the implemented HLLC Riemann solver on the moving mesh is robust for simulating relativistic outflow.

\subsection{Relativistic jet emerged from a black hole-torus system}

\begin{figure*}[!htb]
\centering
\subfloat{
\includegraphics[width=1.0\textwidth]{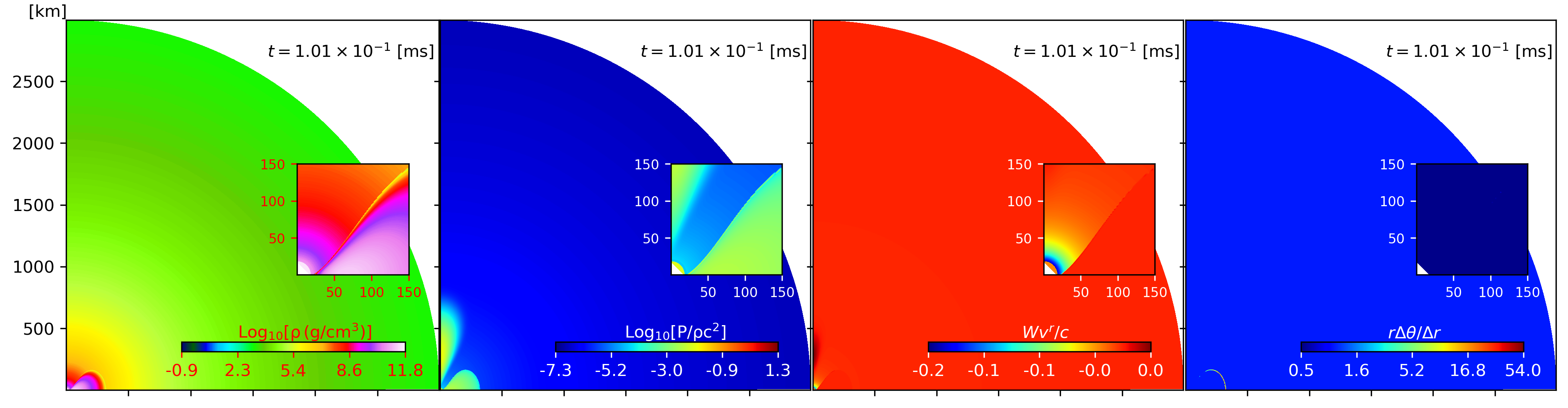}
\label{fig:subfig1}
}\\[-2ex]
\subfloat{
\includegraphics[width=1.0\textwidth]{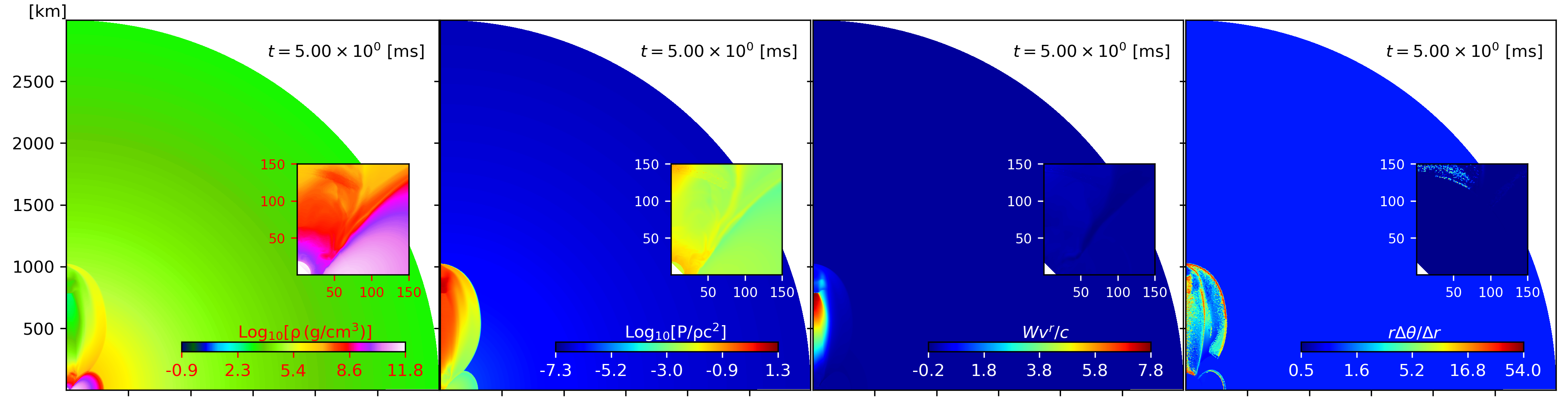}
\label{fig:subfig2}
}\\[-2ex]
\subfloat{
\includegraphics[width=1.0\textwidth]{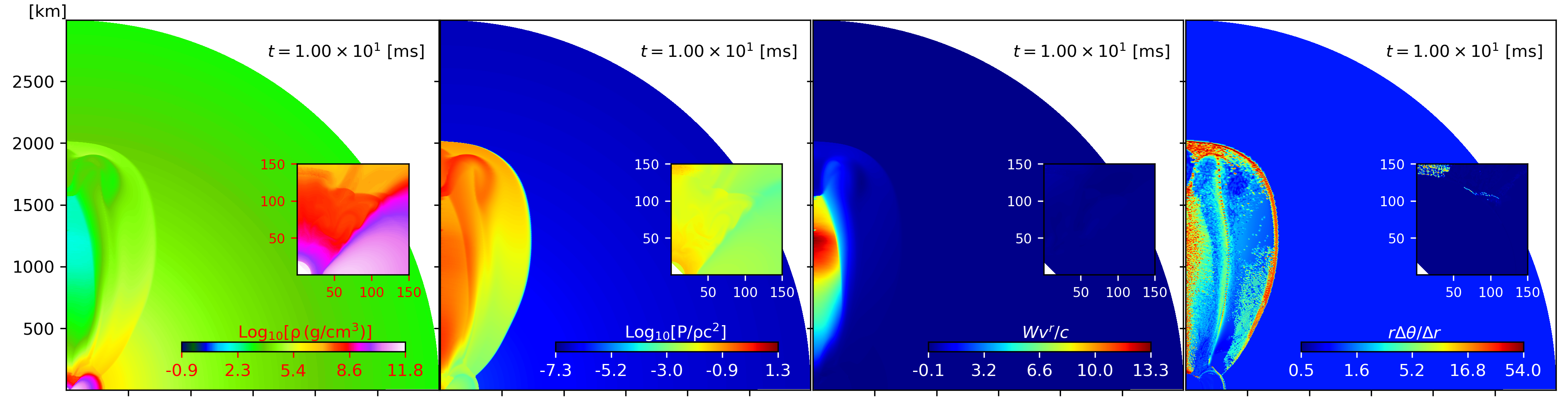}
\label{fig:subfig3}
}\\[-2ex]
\subfloat{
\includegraphics[width=1.0\textwidth]{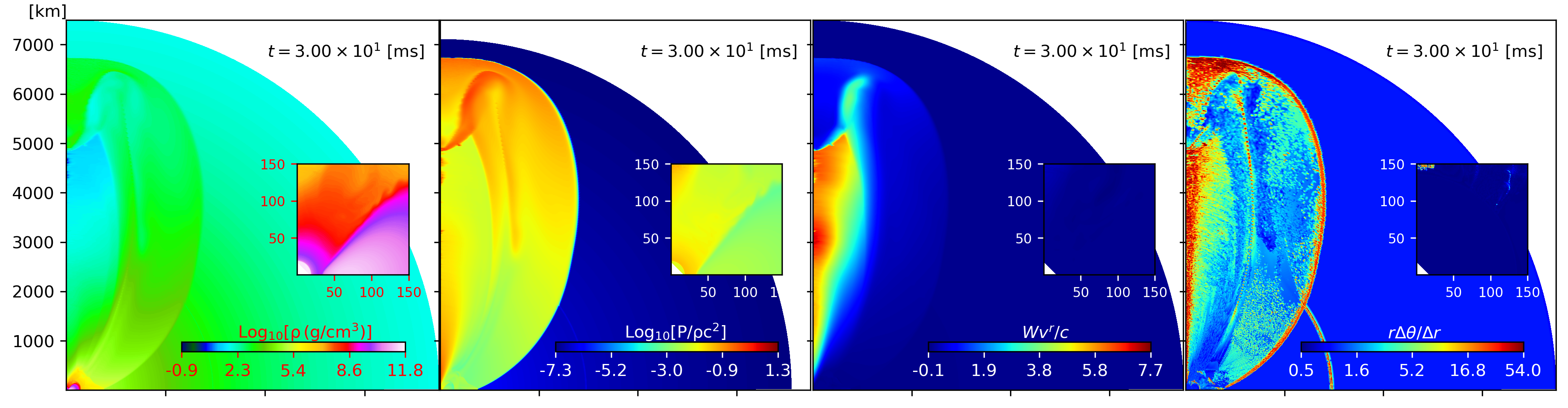}
\label{fig:subfig4}
}
\caption{Jet launching process from a black hole-torus system, visualized at four-time snapshots. From left to right, the contour plots represent the logarithmic density, temperature-like variable $\Theta$, normalized radial velocity $W v^r/c$, and the cell's radial resolution (i.e. inverse aspect ratio). The inner contour plots zoom in on the central region of the domain. The inner boundary is stationary before $t=30\,\rm{[ms]}$. The simulation video is available from Youtube at \cite{videoJet} with high-definition  video quality available.} \label{fig:torusjet_early}
\end{figure*}

\begin{figure*}[h]
\centering
\subfloat{
\includegraphics[width=1.0\textwidth]{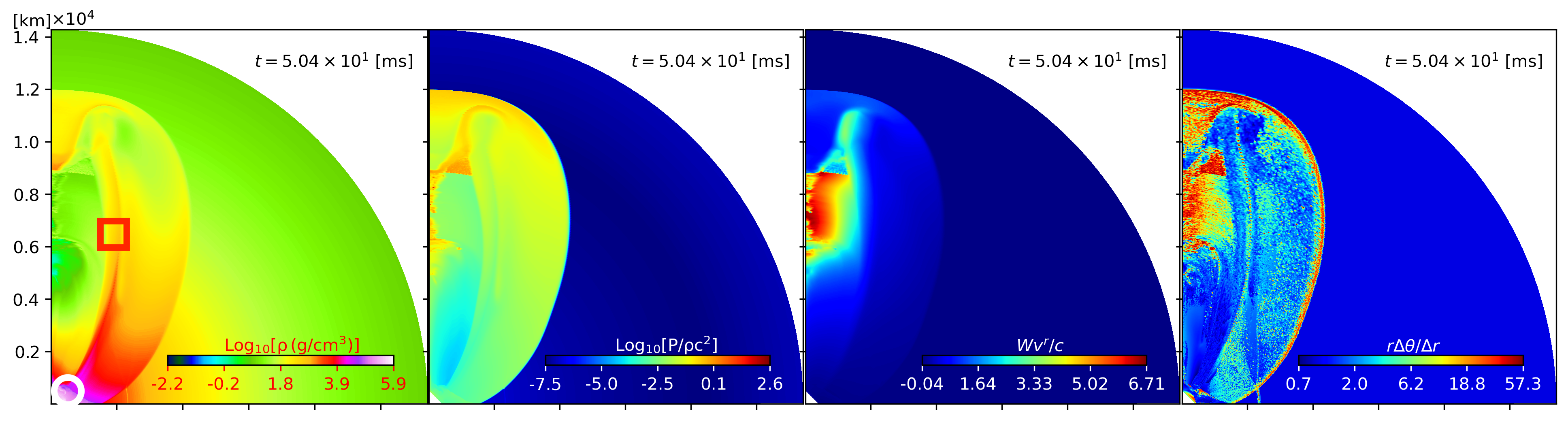}
\label{fig:late_subfig1}
}\\[-2ex]
\subfloat{
\includegraphics[width=1.0\textwidth]{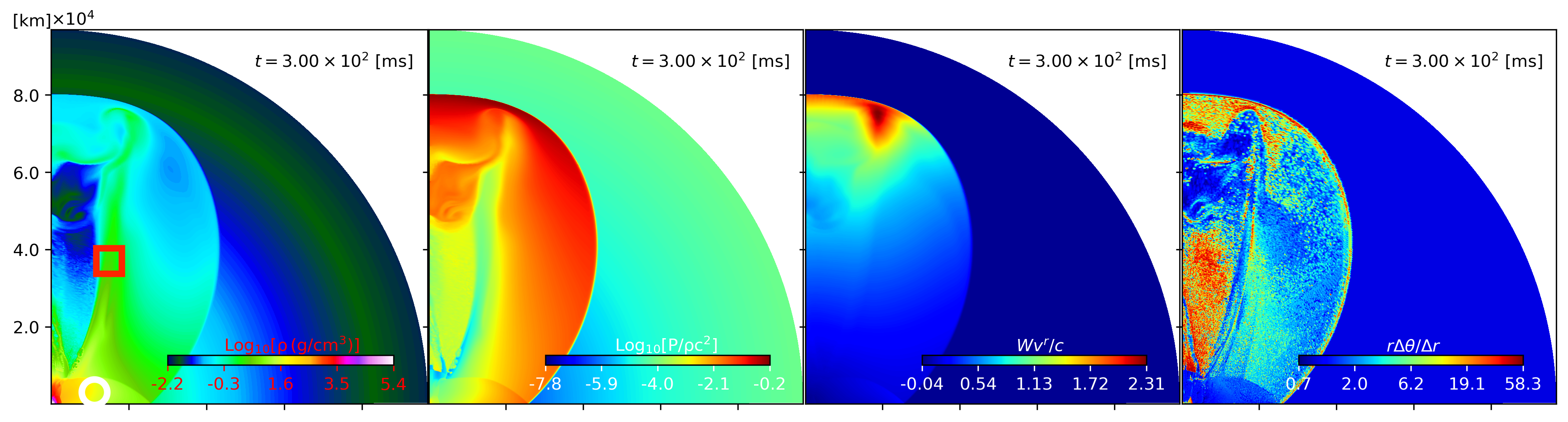}
\label{fig:late_subfig2}
}\\[-2ex]
\subfloat{
\includegraphics[width=1.0\textwidth]{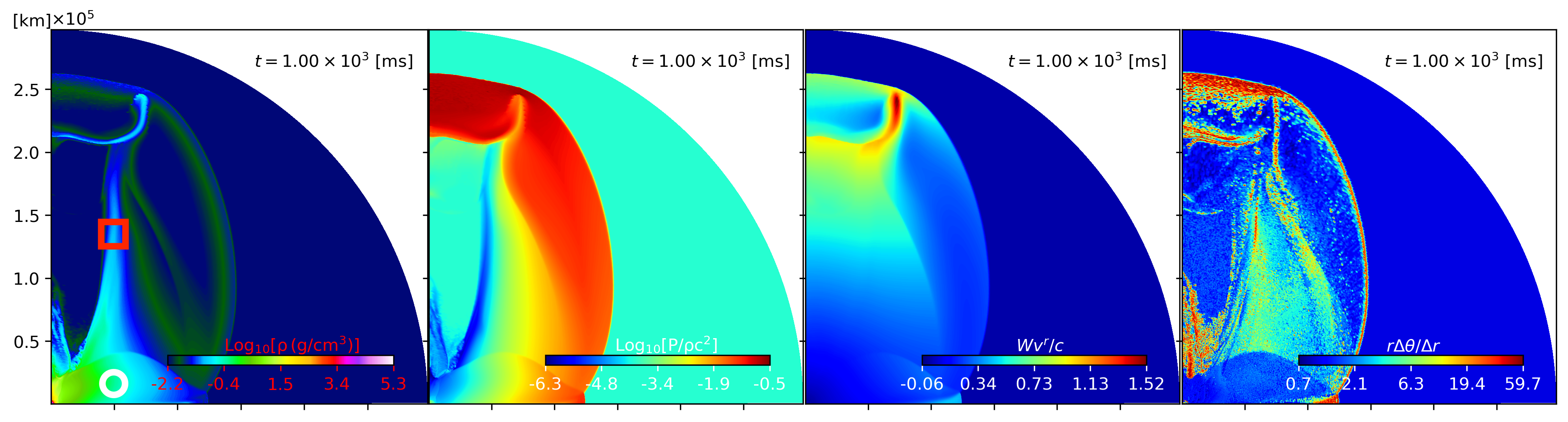}
\label{fig:late_subfig3}
}\\[-2ex]
\subfloat{
\includegraphics[width=1.0\textwidth]{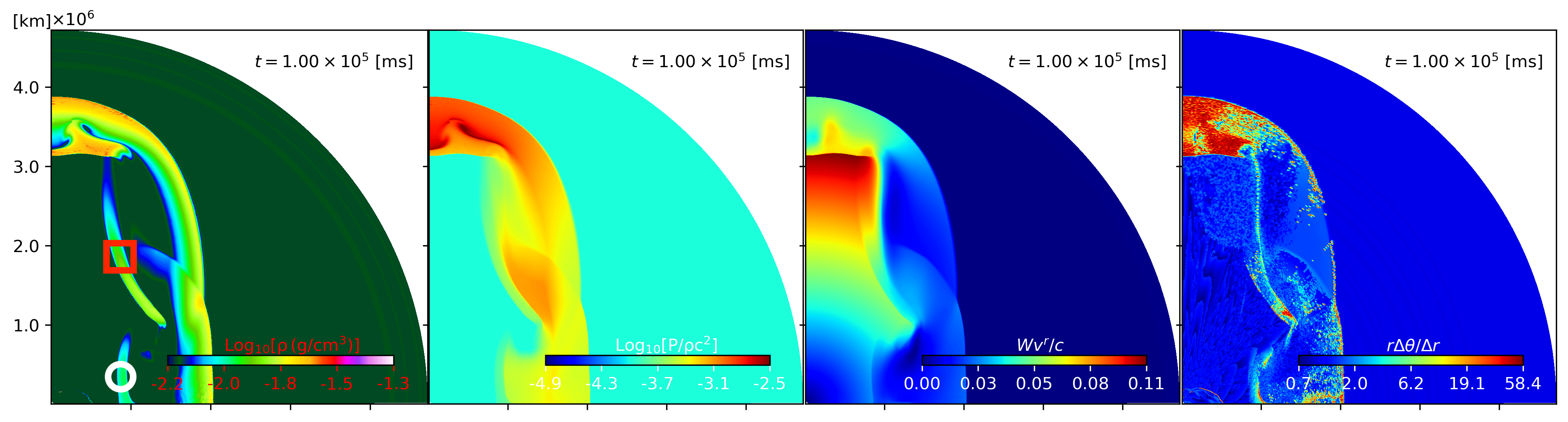}
\label{fig:late_subfig4}
}\\[-2ex]
\subfloat{
\includegraphics[width=1.0\textwidth]{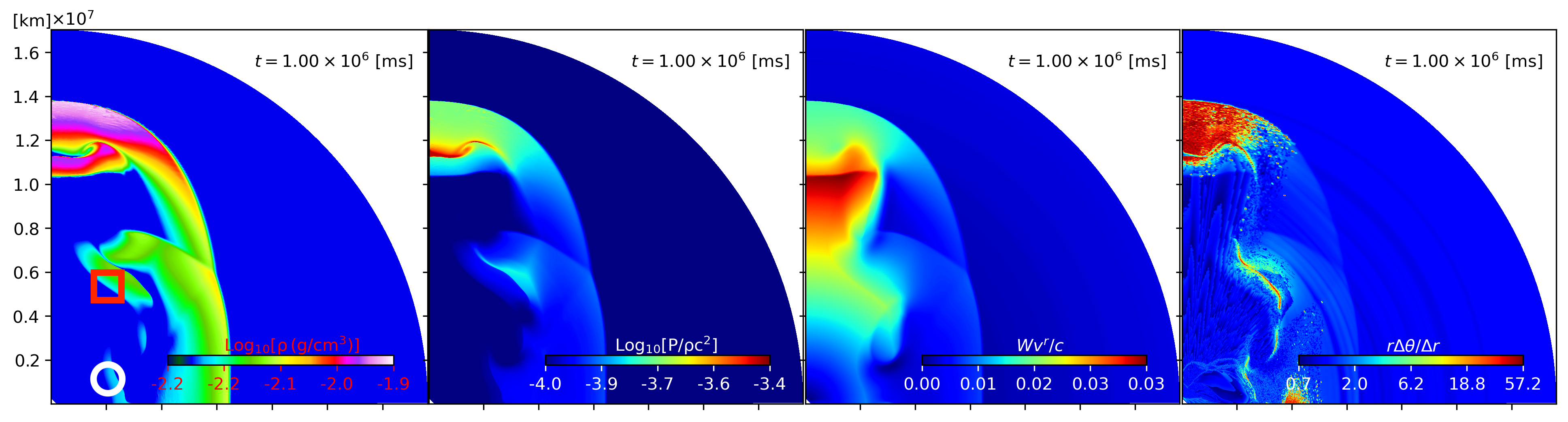}
\label{fig:late_subfig5}
}
\caption{The full-time-domain evolution of the relativistic jet emerged from a black hole-torus system. The inner boundary moves outward at a fraction of the local maximum velocity. The simulation video is available from Youtube at \cite{videoJet} with high-definition video quality available.} \label{fig:torusjet_late}
\end{figure*}

The detection of the gravitational wave (GW) signal GW170817, coupled with the observations of its electromagnetic (EM) counterpart signifies the commencement of the multimessenger astronomy era \cite{2017ApJ...848L..12A}. Research has demonstrated that the structure of the emerged relativistic outflows plays a crucial role in shaping the afterglow emission in GRB170817A \cite{2018ApJ...856L..18M,2018ApJ...863...58X,2018PhRvL.120x1103L,2018ApJ...863L..18A,2018Natur.561..355M,2019MNRAS.489.1919T,2019Sci...363..968G,2019ApJ...870L..15L}. This event provides an ideal candidate for utilizing the electromagnetic observations of the emerged outflow to infer the BNS merging physics. While the presented moving-mesh code is capable of simulating relativistic jets out of various progenitor systems, in the following, we will use a pseudomodel inspired by the outcome of compact binary merger simulations (see e.g., \cite{2014ApJ...784L..28N,2015PhRvD..92b4014K,2017CQGra..34j5014D,2018ApJ...869..130R,2019MNRAS.489L..91C,PhysRevD.101.103002}). We set up a black hole-torus system in the isotropic coordinate of Schwarzschild metric, with the mass of the central black hole $M$ having been set to $M=3 M_{\odot}$ and the torus mass set to $0.2\,M_{\odot}$. The radius of the inner edge of torus is $6\,M$, and the radius of its pressure maximum is set to $16\,M$ \cite{2004ApJ...611..977M}. For the simulation domain, the radius of the inner boundary locates at $12$. And we use $\texttt{Nt}=256$ grids to cover the half spherical domain with cell's initial aspect ratio been set to 1. We adopt the reflecting boundary condition in the azimuthal direction. Outside the torus, the domain is filled with an ejecta cloud with a total mass of $0.02\,M_{\odot}$. The cloud density structure follows:
\begin{equation}
  \rho = \rho_0 (r/r_0)^{-n} + \rho_{\rm{floor}}.
\end{equation}
$\rho_0= (n-3) M_{\rm{ejecta}}/(4\pi r_0^3)$ is derived to give a total ejecta mass $M_{\rm{ejecta}}$. The pressure is $P=K \rho^{\Gamma}+P_{\rm{floor}}, K= 0.54,\Gamma=4/3$. We also add a density floor $\rho_{\rm{floor}}=6.2\times 10^{-5}\,\rm{[g/cm^3]} (\,\rm{i.e.}\, 10^{-22}\rm{\ in\ the\ code\ unit})$ and a pressure floor $P_{\rm{floor}}=10^{-4}\rho_{\rm{floor}}$ to the initial profile to avoid numerical precision error. We set the density slope index $n=3.5$ to represent the postmerger ejecta profile. Here, we ignore the ejecta profile velocity for simplicity. The reference radius $r_0$ is set to $6 M_{\odot}$. A jet engine with a variable luminosity of $L_0=2\times 10^{51}\,\rm{exp}(-t/t_{\rm{decay}})\,[\rm{erg/s}]$ operates for $30\,\rm{ms}$, in the polar region just above the black hole-torus plane. The engine decay timescale $t_{\rm{decay}}$ has been set to $100\,\rm{ms}$. This gives a total injected jet engine energy $5.2\times 10^{49}\,[\rm{erg}]$. We choose this low-energy jet engine injection to test the code's capability of launching a relativistic jet under constraint. In jet simulations, it becomes easier to successfully launch a relativistic jet given a higher energy injection (see e.g. \cite{2018ApJ...863...58X,2019ApJ...880..135X}). The profile of the jet engine features a narrow nozzle with an opening angle of $0.1\,\rm{rad}$. For the complete jet engine profile, we refer readers to the description in Appendix \ref{append:engine_source} as well as in \cite{2019ApJ...880..135X,2015ApJ...806..205D}.

Figure~\ref{fig:torusjet_early} shows the jet launching process during the first $30\,\rm{ms}$. At the beginning of the simulation, the cloud flows into the black hole. In the polar direction, at a location centered around $130\,[\rm{km}]$, a small amount of relativistic gas with a terminal Lorentz factor 100 (i.e. jet engine) gets injected into the cloud. The injected gas has an initial boost velocity in the radial direction (see Appendix \ref{append:engine_source}). The addition of the relativistic gas slightly pushes the cloud gas in the polar direction, leading to a non-negative radial velocity (as can be seen from the radial velocity plot at $t=0.1\,[\rm{ms}]$). The continuous injection of hot relativistic gas drives shocks and changes the temperature profile in the polar direction. By the time $t=5\,[\rm{ms}]$, a shocked cocoon develops and reveals a two-layer structure: a high-density layer which results from the forward shock, meanwhile the inner cocoon which heats up by the jet engine and reverse shock gets to a low-density regime \cite{2011ApJ...740..100B,2018MNRAS.477.2128H,2022ApJ...933L...9G}. Inside the inner cocoon, the shocked gas accelerates to a high velocity with a maximum Lorentz factor around 8 at $t=5\,[\rm{ms}]$. The moving-mesh scheme dynamically allocates cells to resolve the shocked region. The interfaces of the double-layer structure can be seen in the contour plot for the cell's radial resolution: the first interface lies in the shock front between the cocoon and the unperturbed cloud, the second interface is between the cocoon's inner low-density hot relativistic core and its high-density colder part. At the bottom of the cocoon, the shock front hits the torus. At $t=10\,[\rm{ms}]$, the shock front starts to move beyond the torus and wrap around it. At the head of the cocoon, the loaded matter diverts part of the shocked gas sideways. Below this region, the inner core of the cocoon accelerates to a higher Lorentz factor of 13. Throughout the acceleration period, the maximum Lorentz factor of the jet reaches $20$ (which happens at about $t=13\,[\rm{ms}]$), smaller than the terminal Lorentz factor of the injected relativistic gas. This is largely due to the engine's relative low-energy budget (we refer readers to more energetic jet simulations in \cite{2018ApJ...863...58X,2019ApJ...880..135X}).  By the end of the jet engine injection $t=30\,[\rm{ms}]$, at the base of the grid domain, the frontier of the shocked cocoon has passed the torus region. A relative high-density buffer zone appears between the torus and the cocoon (see the density and temperature contour plots). The torus itself rotates stably during the jet launching process, as illustrated by the inner contour plots in Fig.~\ref{fig:torusjet_early}. The head of the shocked cocoon expands beyond the initial grid domain boundary. More cells will be allocated in front of the boundary as the shock front propagates. The radius of the new boundary will make sure that the head of the shock front will stay below 0.8 of this new radius during the simulation.

Figure~\ref{fig:torusjet_late} illustrates the continued evolution of the relativistic cocoon that emerged from the black hole-torus system when the jet engine had been turned off. For computational efficiency, we cut the grid domain within a radius of $300$ (around $440\,[\rm{km}]$). We let the inner boundary move with a velocity of a fraction of the fluid's local maximum velocity. When the inner and outer boundary expands outward together, the simulation gets into a (weak) scaling region where the simulation time step $\Delta t$ increases with the absolute time itself. In this way, the moving-mesh code in spherical coordinates can simulate long-term evolution of relativistic jets over multiple orders of magnitude of time. When the jet engine turns off, it stops accelerating nearby gas. The inner cocoon turns into a shrinking relativistic bubble. The relativistic bubble keeps pushing against the mass-loaded head, exhausting its internal energy and kinetic energy. By the time $t=50\,[\rm{ms}]$, the top of the bubble starts to decelerate dramatically. The collision with the mass-loaded head converts part of its kinetic energy into thermal energy. The collision drives a wave passing through the relativistic bubble, increasing the temperature of inner cocoon all the way to its bottom (see the simulation video). Eventually, the relativistic bubble turns into a relativistic thin shell (see the velocity contour at $t=300\,[\rm{ms}]$). The relativistic shell features a relativistic core with a mildly relativistic sheath, similar to previous special relativistic jet simulations (see, e.g., \cite{2014ApJ...784L..28N,2017ApJ...848L...6L,2018ApJ...863...58X,2019ApJ...880..135X}). The outer layer of the cocoon goes through adiabatic expansion. The density within this layer keeps decreasing while the temperature structure roughly remains the same (see the contour plots for the temperaturelike variable from $t=50\,\rm{[ms]}$ to $t=1\,\rm{[s]}$). The interface between the inner and outer cocoon features a high-density pillar. The relativistic shell keeps sweeping through the medium while depleting its kinetic energy. By the time $t=100\,s$, the relativistic thin shell is replaced by a mass-loaded slow-moving core. We see a morphological change in the outer shell structure. Finally, by the end of our simulation $t=1000\,s$, the outflow velocity becomes completely Newtonian. Now we have seen the complete life cycle of a relativistic jet from its birth at a black hole scale to the distance of its dissipation. In the following, we would like to discuss two dynamical features for this specific simulation. The first feature of interest focuses on the base of cocoon. In Fig.~\ref{fig:torusjet_late}, we use white circle to indicate this region of interest. We find it appears after the shock front of the cocoon passed over the torus. It originates from the buffer zone or shock zone between the original torus and the remaining cocoon. It propogates subrelativistically and maintains its hump shape before $t=1\,\rm{[s]}$. Later on, the shock front accumulates enough matter and slows down to Newtonian velocities. When this happens, via hydrodynamical interaction, the morphology of the region changes and the hump shape disappears, leaving behind a broken filament as shown at $t=100\,s$ and $t=1000\,s$. The second feature of interest relates to the density pillar at the interface between the inner and outer cocoon. We mark it with a red square in the figure. Its formation, to some extent, comes from the shutdown of the central jet engine during the jet launching period. At the beginning, when the central jet engine inflates a cocoon, it drives mass and energy into the cocoon outer layer while creating a low-density hot inner funnel to generate relativistic outflow. When the central engine shuts down, the inner cocoon quickly gets cold and stops pushing the outer layer (see snapshots at $t=50\,\rm{[ms]}$ and $t=300\,\rm{[ms]}$). Then the adiabatic expansion of the outer layer further separates this interface from the shock front as shown at $t=1\,\rm{[s]}$. The interface pillar also has positive radial velocity and moves with the outer shell. However, the part, connecting to the outer shell, moves faster. To a point, the pillar detaches itself from the outer shell and falls back to the inner region. This is what happens from $t=100\,\rm{s}$ to $t=1000\,\rm{s}$. Because of the long-term simulation of the relativistic jet, we are able to capture such detailed hydrodynamics evolution, which may provide insights for the study of morphologies of astronomical jets.  Throughout the simulation, the maximum grid resolution in the radial direction remains below 60--a value we set initially. We see that the moving-mesh scheme, combined with the dynamical grid refinement and derefinement can capture the detailed dynamical features for the relativistic jet simulation over many orders of magnitude of space and time. \added[id=1]{Also the adjusted moving-mesh scheme makes the simulation of relativistic jets computationally efficient. The presented simulation has been performed on a single high-performance computing node with 32 Intel Xeon Gold 6148 CPUs. The whole simulation consumes around 6400 core hours.}

\section{Conclusion} \label{sec:conclusion}

This paper presents an advancement in computational astrophysics: developing and implementing a general relativistic moving-mesh hydrodynamic code featuring an advanced Riemann solver in curvilinear coordinates. 
We showcase the details of integrating a general relativistic framework into the hydrodynamic simulation code \textit{JET}, achieved through applying the reference metric method.

With its ability to elegantly handle the intricate spacetime geometries inherent in general relativity, the tetrad formulation is an ideal choice to address the HLLC Riemann problem under strong gravity.  The achievement of our work is the successful adaptation of the tetrad formulation to incorporate the HLLC Riemann solver into the general relativistic moving-mesh code. We have conducted a series of numerical simulations to validate and demonstrate the efficacy of our newly developed code. These simulations encompass both fixed-mesh and moving-mesh scenarios, allowing us to test the code’s performance under various conditions. The results from these simulations are particularly noteworthy in demonstrating the code’s robustness and reliability in simulating fluid flows under the influence of strong gravitational fields. 

Compared to the fixed-mesh approach, a moving-mesh scheme can increase the time step for fluid regions with high velocity since it removes the limitation imposed by the bulk velocity (see, e.g., \cite{2022MNRAS.510.1315A}). The moving-mesh approach makes the long-term simulation of relativistic jets computationally feasible (see, e.g., \cite{2015ApJ...806..205D,2018ApJ...863...58X,2019ApJ...880..135X}). By extending the \textit{JET} code's capability of handling relativistic jets from an astronomical scale to the scale of a black hole, we have opened new avenues for the full-time-domain simulation of relativistic jets, from their genesis to dissipation. To demonstrate this possibility, we design a representative prototype model which features a torus around a central black hole. A jet is manually launched in the polar direction, near the black hole-torus system. For the underlining jet launching mechanism, we refer readers to Blandford-Payne \cite{1982MNRAS.199..883B} or Blandford-Znajek \cite{1977MNRAS.179..433B} and related dynamo processes (see e.g.\cite{BRANDENBURG20051,2023PhRvL.131a1401K}). In this work, we prescribe an engine profile to imitate the jet launching process. This setup allows us to explore the dynamics of the jet's journey from its origin near the black hole-torus system to its final Newtonian phase. We found multiple new hydrodynamical features from this end-to-end simulation. For the first time, we have been able to simulate the complete life cycle of a relativistic jet, providing insights into the detailed structures of the cocoon and emerged jet over the whole journey. 

Furthermore, these full-time-domain jet simulations enable the joint investigation of various electromagnetic phenomena associated with relativistic outflows. For the case of BNS mergers, the observational phenomena include the kilo-nova emission from the remnant ejecta (see e.g. \cite{1998ApJ...507L..59L,2005astro.ph.10256K,2010MNRAS.406.2650M}), the GRB prompt and afterglow emission (see e.g. \cite{2006RPPh...69.2259M,2014ARA&A..52...43B,2015PhR...561....1K}), and other related processes. By combining our simulations with GRMHD models of jet-launching processes, we will be able to extend the evolution of outflows to distances relevant to long-term electromagnetic radiation observations. This integrative approach aligns perfectly with the era of multimessenger astronomy, allowing for an unprecedented understanding of the underlying physics in jet-launching systems.

While this paper sets the foundational steps in this direction, the complete realization of these ambitious goals remains a pursuit for future research. The potential for further advancements and discoveries in the field is vast, and our work may catalyze the next generation of astrophysical jet simulations, potentially revolutionizing our understanding of relativistic jets and their associated physics.

\section*{Acknowledgements}

Xiaoyi Xie (X.X.) acknowledges the usage of Yamazaki, Sakura clusters at the Max Planck Computing and Data Facility. X.X. thanks Alexis Reboul-Salze, Kyohei Kawaguchi,Kota Hayashi, Masaru Shibata, Sho Fujibayashi, and Takami Kuroda for helpful discussions. X.X. appreciates Kenta Kiuchi for the informative discussion about the tetrad formulation.

\appendix
\section{General relativistic hydrodynamic equations in reference metric formulation \label{append:gr_equations}}

\subsection{\label{sec:level2} The continuity equation}

The covariant divergence of a vector $V^\mu$ gives
\begin{equation}\label{eq:div1}
\nabla_{\mu} V^{\mu}=\frac{1}{\sqrt{-g}} \partial_{\mu}\left(\sqrt{-g} V^{\mu}\right),
\end{equation}
which holds for any metric and its associated covariant derivative. The covariant divergence of a mixed-index second-rank tensor $A_\mu{ }^\nu$, on the other hand, follows (see e.g. \cite{2014PhRvD..89h4043M})
\begin{equation}\label{eq:div2}
    \nabla_{\nu} A_{\mu}{ }^{\nu}=\frac{1}{\sqrt{-g}} \partial_{\nu}\left(\sqrt{-g} A_{\mu}{ }^{\nu}\right)-A_{\xi}{ }^{\nu} \Gamma_{\nu \mu}^{\xi}.
\end{equation}

We will utilize these two rules to derive the general relativistic hydrodynamic equations.

Let us first apply Eq.~(\ref{eq:div1}) to the continuity Eq.~(\ref{eq:conservation_mass})
\begin{equation}\label{cont_eq}
    \begin{aligned}
0 &=\nabla_{\nu}\left(\rho u^{\nu}\right)=\frac{1}{\sqrt{-g}} \partial_{\nu}\left(\sqrt{-g} \rho u^{\nu}\right), \\
&=\frac{1}{\sqrt{-g}}\left(\partial_{0}\left(\sqrt{-g} \rho u^{0}\right)+\partial_{j}\left(\sqrt{-g} \rho u^{j}\right)\right).
\end{aligned}
\end{equation}

The determinant of the spacetime metric leads to
\begin{equation}
\sqrt{- g}=\alpha \sqrt{ \gamma } = \alpha \psi^6 \sqrt{\bar{\gamma}} = \sqrt{\hat{\gamma}} \alpha \psi^6 \sqrt{\frac{\bar{\gamma}}{\hat{\gamma}}}.
\end{equation}

Inserting the previous result into Eq.~(\ref{cont_eq}) we obtain
\begin{equation}
    \partial_{0}\left(\psi^6 \sqrt{\bar{\gamma} / \hat{\gamma}} D\right) + \frac{1}{\sqrt{\hat{\gamma}}}\partial_{j}\left[\sqrt{\hat{\gamma}} \left(f_{D}\right)^{j}\right]=0,
\end{equation}
which is sometimes written as
\begin{equation}
    \partial_{0}\left(\psi^6 \sqrt{\bar{\gamma} / \hat{\gamma}} D\right) + \hat{\mathcal{D}}_j\left[\left(f_{D}\right)^{j}\right]=0,
\end{equation}
where $\hat{\mathcal{D}}_j$ is the covariant derivative with respect to the reference metric $\hat{\gamma}$.
In the above expression, we have defined the density as seen by a normal observer as
\begin{equation}
D=W\rho,
\end{equation}
and the corresponding flux
\begin{equation}
(f_D)^j = \alpha \psi^6 \sqrt{\bar{\gamma}/\hat{\gamma}} D(v^j - \beta^j/\alpha).
\end{equation}

Note that, the fluid velocity as measured by a normal observer, $v^i$, is given by the ratio between the projection of the four-velocity, $u^{\mu}$, in the space orthogonal to $n^{\mu}$ and the Lorentz factor of $u^{\mu}$ as measured by a normal observer, $W= -n_{\mu}u^{\mu}=\alpha u^t$:
\begin{equation}
v^i= \frac{\gamma^i_\mu u^{\mu}}{\alpha u^t} = \frac{u^i}{W}+\frac{\beta^i}{\alpha}.
\end{equation}

\subsection{The Euler equation}
To derive the Euler equation, we apply Eq.~(\ref{eq:div2}) to the projected conservation of energy momentum (\ref{eq:conservation_energy_momentum})
\begin{equation}\label{euler_eq}
    \begin{aligned}
0 &=\gamma_{i \mu} \nabla_{\nu} T^{\nu \mu}=g_{i \mu} \nabla_{\nu} T^{\nu \mu}=\nabla_{\nu}\left(g_{i \mu} T^{\nu \mu}\right) \\
&=\frac{1}{\sqrt{-g}} \partial_{\nu}\left(\sqrt{-g} T_{i}^{\nu}\right)-T_{\nu}^{\mu} \Gamma_{i \mu}^{\nu}.
    \end{aligned}
\end{equation}

We then get the following Euler equation
\begin{equation}
  \frac{1}{\sqrt{-g}}\left(\partial_{0}\left(\sqrt{-g} T_{i}^{0}\right)+\partial_{j}\left(\sqrt{-g} T_{i}^{j}\right)\right) = T_{\nu}^{\mu} \Gamma_{i \mu}^{\nu}. \label{eq:euler_eq}
\end{equation}
Note that the source term leads to
\begin{equation}
  T^{\mu}_{\nu} \Gamma_{i\mu}^{\nu} = T^{\mu \xi} g_{\xi \nu}\Gamma^{\nu}_{i\mu} =  \frac{1}{2} T^{\mu \xi} g_{\xi \mu,i}.
  \label{eq:euler_source}
\end{equation}

Let us define
\begin{subequations}
\begin{align}
\begin{split}
    T^{\mu\nu} &= \rho h u^{\mu} u^{\nu} + P g^{\mu \nu} \\
    &= S_0 n^{\mu} n^{\nu} + S^{\mu}n^{\nu} + S^{\nu} n^{\mu} + S^{\mu \nu}, 
\end{split} \\
    S_0&= \rho h W^2 - P, \\
    S^i &= \rho h W^2 v^i, \\
    S^{ij} &= \rho h W^2 v^i v^j + P\gamma^{ij},
\end{align}
\end{subequations}
and calculate the source term (\ref{eq:euler_source}). The result is shown below (for the derivation, we refer readers to numerical relativity books \cite{shibata2015numerical})
\begin{align}
\begin{split}
    \frac{1}{2} \sqrt{-g} T^{\mu\nu} g_{\mu\nu,i} &= \sqrt{-g} \left( \frac{1}{2} S^{jk} \partial_i \gamma_{jk} + S^\mu n^{\nu} \partial_i g_{\mu\nu} \right.\\
&\left.+  \frac{1}{2} S_0 n^{\mu} n^{\nu} \partial_i g_{\mu \nu}
\right) \\
&= \sqrt{-g}
    \left(\frac{1}{2} S^{jk} \partial_i \gamma_{jk} + \frac{1}{\alpha}S_j \partial_i \beta^j - S_0 \partial_i \ln \alpha  \right).
\end{split}
\end{align}

Let us define the momentum flux as
\begin{equation}
    \begin{aligned}
\left(f_{S}\right)_{i}{ }^{j} & \equiv \alpha \psi^6 \sqrt{\bar{\gamma} / \hat{\gamma}} T_{i}^{j} \\
&=\alpha \psi^6 \sqrt{\bar{\gamma} / \hat{\gamma}}\left(W^{2} \rho h v_{i}\left(v^{j}-\beta^{j} / \alpha\right)+ P\delta_{i}^{j}\right) .\nonumber
\end{aligned}
\end{equation}

We can then rewrite the Euler equation (\ref{eq:euler_eq}) in the following form
\begin{equation}
    \partial_{0}\left(\psi^6 \sqrt{\bar{\gamma} / \hat{\gamma}} S_{i}\right)+ \frac{1}{\sqrt{\hat{\gamma}}}\partial_{j}\left[ \sqrt{\hat{\gamma}} \left(f_{S}\right)_{i}^{\ j}\right]=\left(s_{S}\right)_{i}.
\end{equation}

The definition of the source term is given accordingly
\begin{equation}
\begin{aligned}
  (s_S)_i &= \frac{1}{2}\alpha \psi^6 \sqrt{\frac{\bar{\gamma}}{\hat{\gamma}}} S^{jk} \partial_i \gamma_{jk} \\
  &+  \psi^6 \sqrt{\frac{\bar{\gamma}}{\hat{\gamma}}}  S_j \partial_i \beta^j - \psi^6\sqrt{\frac{\bar{\gamma}}{\hat{\gamma}}} S_0 \partial_i \alpha,
\end{aligned}
\end{equation}
where the first term can be calculated as
\begin{equation}
\begin{aligned}
S^{jk} \partial_i \gamma_{jk} &= S^{jk} \partial_i \left( \psi^4 \bar{\gamma}_{jk} \right) \\
&= S^{jk} \gamma_{jk} \psi^{-4} \partial_i \psi^4 
+ \psi^4 S^{jk} \partial_i (\hat{\gamma}_{jk} + \epsilon_{jk}) \\
&= (\rho h (W^2-1) + 3P) 4\partial_i \ln \psi \\
&+ \psi^4 S^{jk} \partial_i (\hat{\gamma}_{jk}) + \psi^4 S^{jk}\partial_i(\epsilon_{jk})
\end{aligned}.
\end{equation}

\subsection{The energy equation}
For the energy equation, we consider a projection along the normal $n_{\mu}$ of the conservation of
energy-momentum (\ref{eq:conservation_energy_momentum}) and add the conservation of
rest mass (\ref{eq:conservation_mass})
\begin{equation}
    n_{\nu} \nabla_{\mu} T^{\mu \nu} + \nabla_{\mu}\left(\rho u^{\mu}\right)=0,
\end{equation}
which can be rewritten as
\begin{equation}
    \nabla_{\mu}\left(n_{\nu} T^{\nu \mu}+\rho u^{\mu}\right)=T^{\mu \nu} \nabla_{\mu} n_{\nu}.
\end{equation}

We again evaluate the divergence of a vector on the left-hand side, and proceed exactly the same as for the continuity equation, which leads to
\begin{equation}\label{eq:energy_eq}
    \partial_{0}\left(\psi^6 \sqrt{\bar{\gamma} / \hat{\gamma}} \tau\right)+\frac{1}{\sqrt{\hat{\gamma}}} \partial_j\left[ \sqrt{\hat{\gamma}} \left(f_{\tau}\right)^{j}\right] =-\alpha \psi^6 \sqrt{\bar{\gamma} / \hat{\gamma}} T^{\mu \nu} \nabla_{\mu} n_{\nu},
\end{equation}
where we have defined $\tau$ as the internal energy observed by a normal observer
\begin{equation}
    \tau \equiv n_\mu n_\nu T^{\mu\nu}-D \equiv \rho h W^{2} - P-D,
\end{equation}
and the corresponding flux 
\begin{equation}
    \left(f_{\tau}\right)^{j} \equiv \alpha \psi^6 \sqrt{\bar{\gamma} / \hat{\gamma}}\left(\tau\left(v^{j}-\beta^{j} / \alpha\right)+P v^{j}\right).
\end{equation}

The right-hand side of Eq.~(\ref{eq:energy_eq}) (denoted as $s_{\tau}$) can be derived as \cite{2014PhRvD..89h4043M}
\begin{equation}
    \begin{aligned}
      s_{\tau} &\equiv \alpha \psi^6 \sqrt{\bar{\gamma} / \hat{\gamma}}\left(T^{00}\left(\beta^{i} \beta^{j} K_{i j}-\beta^{i} \partial_{i} \alpha\right)\right.\\
&\left.+T^{0 i}\left(2 \beta^{j} K_{i j}-\partial_{i} \alpha\right)+T^{i j} K_{i j}\right).
\end{aligned}
\end{equation}

Finally, we get the energy equation as
\begin{equation}
    \partial_{t}\left(\psi^6 \sqrt{\bar{\gamma} / \hat{\gamma}} \tau\right)+\frac{1}{\sqrt{\hat{\gamma}}}\partial_j \left[ \sqrt{\hat{\gamma}} (f_{\tau})^j \right]=s_{\tau}.
\end{equation}


\subsection{Special relativistic hydrodynamics equations} \label{app:hydro_eq_sr}
Here we write down the special relativistic hydrodynamics equations in spherical coordinate as a comparison to Eq.~(\ref{eq:complete_eqs})
\begin{widetext}
\begin{subequations}\label{eq:complete_eqs_sr}
\begin{align} 
        \frac{\partial}{\partial t} D 
        &+ \frac{1}{r^2} \frac{\partial}{\partial r}\left[r^2 D v_r\right] 
        + \frac{1}{r \sin \theta} \frac{\partial}{\partial \theta} \left[ \sin \theta D v_\theta \right] 
        +  \frac{1}{r\sin \theta} \frac{\partial}{\partial \phi}\left[ D v_\phi\right] 
        = 0,\\
       \frac{\partial}{\partial t} S_r 
        &+ \frac{1}{r^2}  \frac{\partial}{\partial r} \left[ r^2(S_r v_r + P)\right] 
        + \frac{1}{r \sin \theta} \frac{\partial}{\partial \theta} \left[ \sin \theta S_r v_\theta\right] 
        + \frac{1}{r\sin \theta} \frac{\partial}{\partial \phi} \left[S_r v_\phi\right]
        = \frac{2P}{r} + \frac{\rho h W^2(v_\theta^2 + v_\phi^2)}{r}, \\
        \frac{\partial}{\partial t} (r S_\theta) 
        &+ \frac{1}{r^2} \frac{\partial}{\partial r} \left[r^2 (rS_\theta) v_r \right] 
        + \frac{1}{r\sin \theta} \frac{\partial}{\partial \theta} \left[ \sin \theta\ r (S_\theta v_\theta + P)\right] 
        + \frac{1}{r\sin \theta} \frac{\partial}{\partial \phi} \left[(r S_\theta) v_\phi\right] 
        = \cot \theta P + \rho h W^2 v_\phi^2 \cot \theta, \\
       \frac{\partial}{\partial t} (r \sin \theta S_\phi) 
        &+ \frac{1}{r^2} \frac{\partial}{\partial r} \left[ r^2 (r \sin \theta S_\phi) v_r \right] 
        + \frac{1}{r \sin \theta} \frac{\partial}{\partial \theta} \left[\sin\theta (r \sin \theta S_\phi)v_\theta \right] 
        + \frac{1}{r \sin \theta} \frac{\partial}{\partial \phi} \left[ r \sin \theta (S_\phi v_\phi + P)\right] = 0, \\
        \frac{\partial}{\partial t} \tau
        &+ \frac{1}{r^2} \frac{\partial}{\partial r} \left[r^2 (\tau v_r + P v_r) \right] 
        + \frac{1}{r\sin \theta} \frac{\partial}{\partial \theta} \left[\sin \theta (\tau v_{\theta} + P v_{\theta}) \right] 
        + \frac{1}{r\sin \theta} \frac{\partial}{\partial \phi} \left[ \tau v_{\phi} + P v_{\phi}\right] = 0.
\end{align}
\end{subequations}
\end{widetext}
~\newline


\section{Transformation between Schwarzschild coordinates and maximally sliced trumpet coordinate}\label{appendix:bondi}

Starting from a family of stationary, maximal slicing of the Schwarzschild spacetime
\begin{align}
\begin{split}
    &ds^2 = -\alpha^2 dt^2 + \gamma_{ij}(dx^i+\beta^i dt)(dx^j + \beta^j dt)\\
&= -\left(\alpha^2-\beta_R \beta^R\right) d t^2+2 \beta_R d t d R+f^{-2} d R^2+R^2 d \Omega^2,
\end{split}
\end{align}
with lapse $\alpha = f$, shift vector $\beta^R = \frac{Cf}{R^2}$,
$f= \left(1-\frac{2M}{R} + \frac{C^2}{R^4} \right)^{1/2}$ and $C$ being the integration constant.
The transformation into isotropic coordinates follows
\begin{equation}
f^{-2} d R^2+R^2 d^2 \Omega=\psi^4\left(d r^2+r^2 d^2 \Omega\right),
\end{equation}
so we have $dR/dr = \psi^2 f, R=\psi^2 r$.
Two simple solutions of $r$ and $\psi$ can be found for the cases $C=0$ and $C=3\sqrt{3}M^2/4$. 
For the case of $C=0$, the solution yields the familiar isotropic coordinate of the Schwarzschild metric, 
\begin{align}
\begin{split}
r= & \frac{R-M + \sqrt{R^2 - 2M R}}{2}, \\
\alpha= & \left(1-\frac{2M}{R} \right)^{1/2} = \frac{1-M/(2r)}{1+M/(2r)}.
\end{split}
\end{align}

For the case of $C=3\sqrt{3}M^2/4$, the solution yields the isotropic trumpet coordinate \cite{PhysRevD.75.067502,2017CQGra..34c5007M,2022PhRvD.106l4041K}
\begin{align}
\begin{split}
r= & {\left[\frac{2 R+M+\left(4 R^2+4 M R+3 M^2\right)^{1 / 2}}{4}\right] } \\
& \times\left[\frac{(4+3 \sqrt{2})(2 R-3 M)}{8 R+6 M+3\left(8 R^2+8 M R+6 M^2\right)^{1 / 2}}\right]^{1 / \sqrt{2}}, \\
\alpha= &\left(1-\frac{2 M}{R}+\frac{27 M^4}{16 R^4}\right)^{1 / 2}, \\
K_{rr} &= -\frac{2\psi^4 C}{R}, \\
K_{\theta\theta} &= \frac{K_{\phi\phi}}{\rm{sin}^2\theta} = \frac{C}{R}, \\
\gamma^{i j} &=\psi^{-4} \eta^{i j}=\left(\frac{r}{R}\right)^2 \operatorname{diag}\left(1, r^{-2}, r^{-2} \sin ^{-2} \theta\right).
\end{split}
\end{align}

In this coordinate, we have
\begin{equation}
K \equiv \gamma^{i j} K_{i j}=-\frac{2 C}{R^3}+\frac{C}{R^3}+\frac{C}{R^3}=0,
\end{equation}
which represents a maximal slicing of the Schwarzschild spacetime with limitng surface at $R=3M/2$ \cite{2017CQGra..34c5007M}.

\section{Analytical jet engine model\label{append:engine_source}}

The jet engine model utilizes the  nozzle function  $g(r,\theta)$ as shown in \cite{2015ApJ...806..205D,2019ApJ...880..135X}. We list its expression here,
\begin{equation}
    g(r,\theta) \equiv (r/r_{\rm{jet}}) e^{-(r/r_{\rm{jet}})^2 /2}e^{ (\cos
\theta - 1) /\theta_0^2 }/N_0,
\end{equation}
where $r_{\rm{jet}}$ is the central position for the jet engine injection, $\theta_0$ is the jet engine opening angle
, and $N_0$ is the normalization of $g$ via the integration over $r\in[0,\infty],\theta\in[0,\pi/2]$,
\begin{equation}
  N_0 \equiv 4 \pi r_{\rm{jet}}^3 ( 1 - e^{-1/\theta_0^2} )
  \theta_0^2.
\end{equation}

For the complete list of jet engine parameter value, we refer readers to Table \ref{table:jet_engine}.
\begin{table}[thb]
\centering
\caption{Parameter values \footnote{$L_0$ and $t_{\rm{decay}}$ represent the jet engine power and the jet engine decay timescale, respectively. The duration of the jet engine is given as $t_0$. $\eta_0$ is the jet engine energy-to-mass ratio. $\gamma_0$ is the jet engine initial Lorentz factor. $\theta_0$ is the jet engine opening angle.} for the jet engine.}\label{table:jet_engine}
\begin{tabular}{ p{2cm}|p{2cm}|p{2cm} }
 \hline
 \hline
  $L_{0}\,\rm{[erg\,s^{-1}]}$ & $t_{\rm{decay}}\,\rm{[s]}$  &  $t_{0}\,\rm{[s]}$   \\
 \hline
   $2\times 10^{51}$  &  0.1  & $0.03$    \\
   \hline
   \hline
   $\eta_0$ & $\gamma_0$ & $\theta_{\rm{0}}$   \\
   \hline
    100  & 5 & 0.1 \\
   \hline
\end{tabular}  
\end{table}

We then inject the jet engine into the domain cells by adding mass, momentum, and energy source into the corresponding conserved variables, according to:
\begin{eqnarray}
  S_0 &=&  L_0 e^{-t/t_{\rm{decay}}} g(r,\theta), \\
  \Delta S_D & = & S_0/\eta_0 \mathcal{M} dVdt, \\
  \Delta S_{S_r} & = & S_0 \sqrt{ 1 - 1/\gamma_0^2 } \mathcal{M} dVdt, \\
  \Delta S_{\tau} &=& (S_0 - S_0/\eta_0) \mathcal{M} dVdt,
\end{eqnarray}
where $S_0$ is the injected jet engine energy profile, and the other three variables are the added source terms. $\mathcal{M}$ is the conformal factor coefficient. $dVdt$ represents the cell's spacetime coordinate volume.

\bibliography{ref.bib}
\bibliographystyle{apsrev4-1}

\clearpage
\onecolumngrid
\listofchanges[show=all]

\end{document}